\documentclass[aps,twocolumn,superscriptaddress,showpacs,showkeys,preprintnumbers,amsmath,amssymb,amsfonts,nofootinbib]{revtex4}

\pdfoutput=1

\usepackage{graphicx}
\usepackage{color}
\usepackage{slashed}
\usepackage{tabularx}
\usepackage{hyperref}

\newcommand{\Tr}{\mathrm{Tr}}

\newcommand{\I}{\mathrm{i}}
\newcommand{\be}{\begin{eqnarray}}
\newcommand{\ee}{\end{eqnarray}}

\newcommand{\Nf}{N_{\text{f}}}
\newcommand{\Nc}{N_{\text{c}}}

\newcolumntype{L}[1]{>{\raggedright\arraybackslash}p{#1}} 
\newcolumntype{C}[1]{>{\centering\arraybackslash}p{#1}} 
\newcolumntype{R}[1]{>{\raggedleft\arraybackslash}p{#1}} 

\graphicspath{{.}{figures/}}
\begin{document}

\title{Spectral Functions for the Quark-Meson Model Phase Diagram\\
from the Functional Renormalization Group}

\newcommand{\TUD}{Theoriezentrum, Institut f\"ur Kernphysik,
  Technische Universit\"at Darmstadt, 64289 Darmstadt, Germany}
\newcommand{\RKU}{Institut f\"ur Theoretische Physik,
 Ruprecht-Karls-Universit\"at Heidelberg, 69120 Heidelberg, Germany}
\newcommand{\JLU}{Institut f\"ur Theoretische Physik,
  Justus-Liebig-Universit\"at Giessen, 35392
  Giessen, Germany}
\newcommand{\GSI}{GSI Helmholtzzentrum f\"ur 
  Schwerionenforschung GmbH, 64291 Darmstadt, Germany}

\author{Ralf-Arno Tripolt}\affiliation{\TUD}
\author{Nils Strodthoff}\affiliation{\RKU}
\author{Lorenz von Smekal}\affiliation{\TUD}\affiliation{\JLU}
\author{Jochen Wambach}\affiliation{\TUD}\affiliation{\GSI}

\begin{abstract}
We present a method to obtain spectral functions at finite temperature
and density from the Functional Renormalization Group. Our method is
based on a thermodynamically consistent truncation of the 
flow equations for 2-point functions with analytically continued
frequency components in the originally Euclidean external momenta. 
For the uniqueness of this continuation at finite temperature we 
furthermore implement the physical Baym-Mermin boundary
conditions. We demonstrate the feasibility of the method by
calculating the mesonic spectral functions in the quark-meson model 
along the temperature axis of the phase diagram, and at 
finite quark chemical potential along the fixed-temperature line  
that crosses the critical endpoint of the model. 
\end{abstract}



\maketitle

\section{Introduction}

Dynamic properties of strongly interacting matter are of fundamental interest 
to many theoretical as well as experimental studies in different branches of physics  
\cite{Arnold:2000dr,Kovtun:2004de,Meyer:2008gt,Schafer:2009dj,Taylor:2010,Meyer:2011gj,Shen:2011eg}. 
Spectral functions, for example, encode information on the particle spectrum as 
well as collective excitations and serve as input for the calculation 
of transport coefficients like the shear viscosity
\cite{Aarts:2002cc,Haas:2013hpa}.
However, the calculation of such real-time observables at finite temperature 
and density represents an inherently difficult problem.

On one hand, in the limit of vanishing external momenta, 
in which all information about $n$-point functions is encoded in the 
effective potential, nonperturbative methods are 
required even for equilibrium thermodynamics.
Although mean-field calculations might capture the gross features of 
a phase diagram, quantitative predictions and correct 
descriptions of critical phenomena, e.g.\ by nontrivial critical exponents,
require the proper inclusion of fluctuations. QCD thermodynamics at
vanishing or nearly vanishing baryon density can be studied
systematically with Monte-Carlo
simulations on the lattice. In the region of higher baryon densities,
however, the fermion sign-problem arises as an additional difficulty 
in Lattice QCD \cite{Philipsen:2011zx}. This is one motivation for
complementary approaches based on functional continuum methods which
allow to include finite chemical potentials essentially
without further complications. The particular one used in this work is
the Functional Renormalization Group (FRG), see
\cite{Berges:2000ew,Polonyi:2001se,Pawlowski:2005xe,Schaefer:2006sr,Kopietz2010,Braun:2011pp,Gies2012}
for reviews.  

On the other hand, a technical difficulty that arises already in the 
vacuum and which is common to all Euclidean 
approaches to Quantum Field Theory is the need to analytically 
continue from imaginary to real time for dynamic processes especially with
timelike momentum transfer. This is even more so at finite temperature
where these continuations are based on data at discrete Matsubara frequencies
and hence require additional boundary conditions for uniqueness
\cite{Baym1961,Landsman1987}. Even with the analytic structure
completely fixed, however, the reconstruction of spectral functions
from discrete numerical (i.e., noisy) data on Euclidean
correlation functions, for example, is an ill-posed inverse
problem. Maximum entropy methods (MEM)
\cite{Jarrell:1996,Asakawa:2000tr}, Pad\'{e} 
approximants \cite{Vidberg:1977}, or very recently also a standard 
Tikhonov regularization \cite{Dudal:2013yva} have been proposed to
deal with this problem. They all work best at low temperatures when
the density of Matsubara modes is sufficiently large, but they all
break down when the Euclidean input data is not sufficiently dense and
precise.  

Therefore any approach that can deal with the analytic 
continuation explicitly is highly desirable. Such alternative approaches
were proposed in \cite{Strodthoff:2011tz, Kamikado2013} and
\cite{Floerchinger2012} to involve an analytic continuation on the
level of the flow equations themselves in order to provide correlation
functions for timelike external momenta as the output of the
calculations. In addition to its simplicity our approach enjoys a
number of particular advantages: First of all, it is thermodynamically
consistent in that the spacelike limit of zero external momentum in
the 2-point correlation functions agrees with the curvature or
screening masses as extracted from the thermodynamic grand potential
\cite{Strodthoff:2011tz}. Secondly, it satisfies the
physical Baym-Mermin boundary conditions  at finite temperature 
\cite{Baym1961} whose implementation here proceeds essentially as in a 
simple one-loop calculation \cite{Das:1997gg}. Finally, although we
focus on mesonic spectral functions in this work it can be extended to
calculate also quark and gluonic spectral functions as an alternative
to analytically continued Dyson-Schwinger equations (DSEs)
\cite{Strauss:2012dg}, or to using MEM on Euclidean FRG
\cite{Haas:2013hpa}  or DSE results
\cite{Nickel:2006mm,Mueller:2010ah,Qin:2013ufa}.  

A further extension would be to feed the full momentum dependence of
the resulting 2-point functions back into the flow equation for the
grand potential to solve this flow equation in an extended truncation,
beyond the leading order derivative expansion. This would yield new 
$3$ and $4$-point couplings to be used in the flow equations for the 2-point
functions again, and this procedure could in principle be iterated to extend the
consistency of the approach to their full momentum-dependence in this way. 

Only such an iterative procedure will
eventually provide a non-trivial width for resonance peaks in
spectral functions. In this work we focus on the first iteration step
of this procedure in which the momentum-dependent 2-point functions
are calculated on the basis of a given solution for the effective
potential. Although the spacelike zero-momentum limits of the 2-point
functions remain fixed to the effective potential, this is obviously
not a fully self-consistent procedure for non-zero external momenta, and
hence for the spectral functions at this point yet. One should keep in
mind, however, that our calculations involve no prior knowledge of the
momentum dependence or assumptions on the singularity structure of the 
propagators that one obtains. 

In extension of previous results for the $O(4)$ linear-sigma 
model in the vacuum \cite{Kamikado2013a}, here we calculate 
mesonic spectral functions at finite temperature and with
dynamical quarks at zero and finite density. This is done within the
quark-meson model, which serves as a low  energy effective model for
QCD with $\Nf=2$ light quark flavors \cite{Jungnickel:1995fp,
  Schaefer:2004en}.
In particular, it realizes the chiral symmetry 
breaking pattern of 2-flavor QCD and allows going beyond the
mean-field approximation by including the fluctuations due to
collective mesonic excitations. It has frequently been employed in
model studies of the QCD phase diagram at finite temperature and
density as a simple and intuitive replacement for more directly QCD
based calculations 
\cite{Braun:2009gm,Fischer:2011mz,Fischer:2012vc,Luecker:2013oda},
and it can systematically be improved towards full QCD 
by using appropriate input to model the influence of the gauge-field dynamics
\cite{Haas:2013qwp,Herbst:2013ufa}.
For parameters that yield realistic quark and meson masses in the vacuum, 
the phase diagram of this model exhibits a first-order phase transition ending 
in a critical point and a crossover. It therefore allows for the 
study of critical regimes in the phase diagram as well as effects of chiral 
symmetry restoration.

This paper is organized as follows. In Sec.~\ref{sec:setup} we briefly
introduce the FRG approach, especially for the quark-meson model, and
discuss our analytic continuation procedure. Results on thermodynamic
observables, the phase diagram of the quark-meson model and mesonic
spectral functions at finite  temperature and chemical potential are
shown in Sec.~\ref{sec:results}. Finally, our summary and a brief
outlook are provided in Sec.~\ref{sec:summary}.

\section{Theoretical Setup}\label{sec:setup}
In this section we introduce the Functional Renormalization Group and
its application to the quark-meson model as a chiral effective model  
for QCD. We derive the flow equations for the scale-dependent
effective potential, and the corresponding ones for the pion and sigma
meson 2-point functions together with our analytic continuation
procedure for the originally discrete Matsubara frequencies as the
zero-components of their external momenta at finite temperature. 

\subsection{Functional Renormalization Group Approach to the Quark-Meson Model}
\label{sec:FRG_QM}
\begin{figure}[t]
\includegraphics[width=\columnwidth]{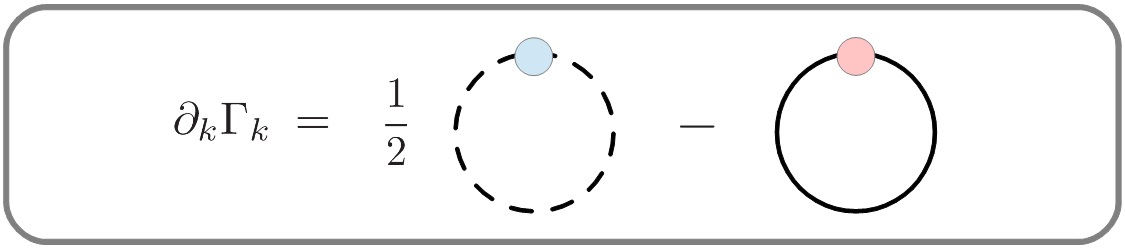}
\caption{(color online) Diagrammatic representation of 
the flow equation for the effective action. Dashed (solid) lines 
represent bosonic (fermionic) propagators and circles represent regulator 
insertions $\partial_k R_k$.}
\label{fig:flow_Gamma} 
\end{figure}

The FRG is a powerful tool for
nonperturbative calculations in quantum field theory and statistical
physics, see  
\cite{Berges:2000ew,Polonyi:2001se,Pawlowski:2005xe,Schaefer:2006sr,Kopietz2010,Braun:2011pp,Gies2012}. 
It involves introducing an infrared (IR) regulator 
$R_k$ to suppress fluctuations from momentum modes with momenta below the associated
renormalization group (RG) scale $k$. This regulator has then to be
removed by taking $k$ from the ultraviolet (UV)
cutoff scale $\Lambda$ down to zero. 
The central object in the approach pioneered by Wetterich
\cite{Wetterich:1992yh} is the  scale-dependent effective average
action $\Gamma_k$ which thereby interpolates between the microscopic
bare action  at $k=\Lambda$ and  the full quantum effective action for
$k\to 0$.  Its scale derivative is governed by an exact one-loop equation
involving full scale- and field-dependent propagators which takes the
form 
\begin{equation}
\label{eq:floweffaction}
\partial_k \Gamma_k=\tfrac{1}{2}\text{STr}{[} \partial_k R_k(\Gamma^{(2)}_k+R_k)^{-1}{]},
\end{equation}
where $\Gamma^{(2)}_k$ denotes the second functional derivative of the
effective average action, and the supertrace includes internal and
spacetime indices, the functional trace, typically in  momentum space,
and the minus sign with degeneracy factor for fermion loops. The flow
equation for the effective average action is represented
diagrammatically for bosonic and fermionic fields  
in Fig.~\ref{fig:flow_Gamma}.

The calculation now proceeds by specifying an Ansatz for the effective
average action  $\Gamma_k$. For the quark meson model in the zeroth
order derivative expansion, where only the effective potential carries
a scale dependence, it reads
\begin{eqnarray} 
\Gamma_{k}[\bar \psi,\psi,\phi]&=& \int d^{4}x \Big\{
\bar{\psi} \left({\partial}\!\!\!\slash +
h(\sigma+i\vec{\tau}\cdot\vec{\pi}\gamma_{5}) -\mu \gamma_0 \right)\psi\nonumber \\
&& \qquad\qquad
+\tfrac{1}{2} (\partial_{\mu}\phi)^{2}+U_{k}(\phi^2) - c \sigma 
\Big\}\,,
\label{eq:QM}
\end{eqnarray}
where $\phi_i=(\sigma,\vec \pi)_i$ and $\phi^2=\sigma^2+\vec
\pi^2$. The explicit symmetry breaking term $c\sigma$ accounts for a
non-vanishing pion mass but does not enter the flow equation and is
consequently only added to the full field-dependent potential at the
IR scale.  

\begin{figure}[ht]
\includegraphics[width=\columnwidth]{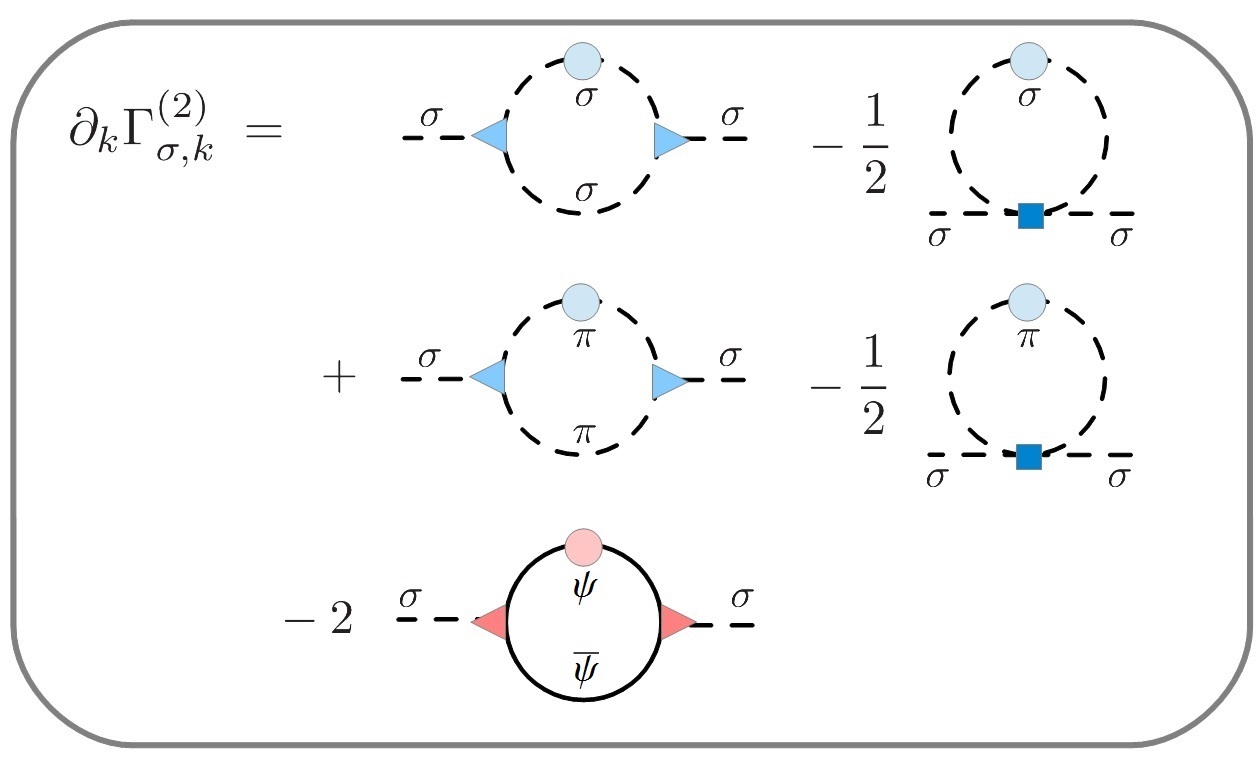}
\includegraphics[width=\columnwidth]{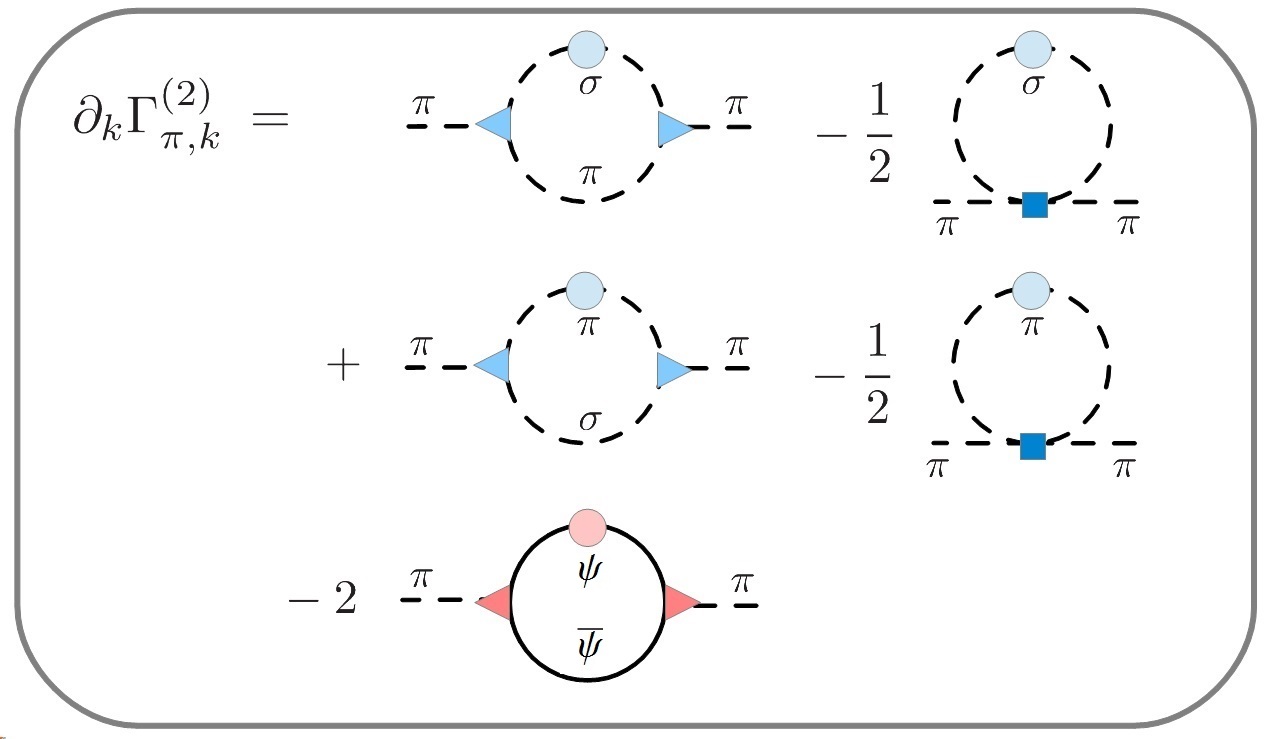}
\caption{(color online) Diagrammatic representation of 
the flow equation for the mesonic 2-point functions. 
Three-point vertices are represented by triangles,
four-point vertices by squares.}
\label{fig:flow_Gamma2} 
\end{figure}

Given the flow equation in Eq.~(\ref{eq:floweffaction}), 
one straightforwardly obtains flow 
equations for the 1-particle irreducible $n$-point functions upon taking $n$
functional derivatives of the flow equation for the effective average
action with respect to the fields. In this work we are particularly
interested in the flow equations for the pion and sigma meson
2-point functions which are 
represented diagrammatically in Fig.~\ref{fig:flow_Gamma2}. The flow
equations for $n$-point functions naturally involve up to
$(n+2)$-point vertices as input which leads to an infinite tower of
coupled equations. Hence, the numerical solution of such a  
system requires the use of truncations. In the following we neglect the 
momentum dependences of higher $n$-point vertex functions, which can
be taken into account by more involved truncation schemes such as the
BMW approximation \cite{Blaizot:2005wd,Blaizot:2006vr} for example,  
and treat them as the RG scale $k$ dependent couplings which we
extract from the flowing effective potential. For now, the input 
2-point functions in the loops on the right are also those obtained
from the scale dependent average action at the leading-order in
the derivative expansion in Eq.~(\ref{eq:QM}). They are thus of
tree-level form but contain the scale-dependent screening masses from
the flow of the effective potential. We emphasize again that this
scheme can be upgraded in the future to iteratively include the nontrivial
momentum-dependences of the resulting 2-point functions as described
in the introduction. The quark-meson 3-point
vertices in the flow equations for the 2-point
functions remain scale-independent at this level,   
\begin{equation}
\label{eq:quark_vertex}
\Gamma^{(2,1)}_{\bar \psi \psi  \phi_i}=h \begin{cases}1 &\text{for}\, i=0\\
\I\gamma^5\tau^i &\text{for}\, i=1,2,3 \end{cases}\,,
\end{equation}
while the mesonic $3$-- and $4$-point vertices are obtained in our
truncation as
\begin{eqnarray}
\label{eq:meson_vertex_3}
\Gamma^{(0,3)}_{\phi_i\phi_j\phi_m}&=& 4 U''_k \left(  \delta_{ij}\phi_m + 
\delta_{im}\phi_j + \delta_{jm}\phi_i \right)\nonumber\\
&&+\,8U^{(3)}_k \phi_i\phi_j\phi_m
\,,\\[3mm]
\label{eq:meson_vertex_4}
\Gamma^{(0,4)}_{\phi_i\phi_j\phi_m\phi_n}&=&4 U''_k 
\left(  \delta_{ij}\delta_{mn} + \delta_{in}\delta_{jm} + \delta_{jn}\delta_{im} 
\right) \nonumber\\
&&+\,
8U^{(3)}_k \left( \delta_{ij}\phi_m\phi_n + \delta_{jm}\phi_i\phi_n + 
\delta_{mn}\phi_i\phi_j \right. \nonumber \\
&& \left. \qquad \quad + \,\delta_{jn}\phi_i\phi_m + 
\delta_{in}\phi_j\phi_m + \delta_{im}\phi_j\phi_n  \right) \nonumber\\
&&+\,
16 U^{(4)}_k \phi_i\phi_j\phi_m\phi_n ,
\end{eqnarray}
where $U^{(n)}_k$ here denotes the $n$th derivative of the effective 
potential with respect to $\phi^2 = \sigma^2 + \vec \pi^2$. 

The determination of higher $n$-point vertices using solely the input 
from the scale-dependent effective potential has the particular advantage 
that the flow equations for the 2-point functions 
evaluated at vanishing external momentum automatically reduce to the
corresponding combinations of derivatives of the flow equation for the
effective potential, as they must in a consistent truncation scheme
based on a single generating object, which is in our case the
effective potential. More explicitly, this implies that
\begin{align}
\partial_k \Gamma^{(0,2)}_{\pi,k} (p=0)&=2 \partial_k U_k'\,, \label{eq:tdcons}\\
\partial_k \Gamma^{(0,2)}_{\sigma,k} (p=0)&=2 \partial_k U_k'+4 
\partial_k U_k''\phi^2\,. \label{eq:tdcons2}
\end{align}
At finite temperature special care has to be taken with respect to the 
order of limits here. Once the frequency component $p_0$ of the external 
momentum is analytically continued, the static limit $\vec p \to 0$ with
$p_0=0$ and the long-wavelength limit $p_0\to 0 $ for $\vec p=0$ do
not commute  \cite{Das:1997gg,Lebellac:1996,Strodthoff:2011tz}. The 
identities in (\ref{eq:tdcons}) and (\ref{eq:tdcons2}) above are then
understood as the spacelike static limit $\vec p\to 0$ of the lowest
Matsubara mode $p_0=0$ before analytic continuation as they determine
the static screening masses. 

In this work we present the first step of the iteration procedure described 
above in which the 2-point functions are obtained by integrating their
corresponding flow equations, see 
Fig.~\ref{fig:flow_Gamma2}, built on the given solution for the effective 
potential as fixed input. While full self-consistency can only be
achieved by the iteration process, already the results for the 2-point 
functions after this first step show a non-trivial momentum
dependence which is not restricted by any Ansatz. Note that
unlike the iteration procedure for the effective potential, which is
self-consistent but requires only 2-point functions evaluated at
Euclidean and hence spacelike external momenta as input, the
calculations of real-time quantities such as spectral functions
furthermore require 2-point functions evaluated at timelike external
momenta which are obtained by analytic continuation as described below.

In the following we present the flow equations for the effective  
potential and the pion and sigma meson 2-point functions. The
flow equation for the effective potential is obtained by evaluating
Eq.~(\ref{eq:floweffaction}) for constant fields and reads, with $N=4$
in our 2-flavor $O(4)$ case, 
\begin{equation}
\label{eq:flow_pot} 
\partial_k U_k =
\tfrac{1}{2} I_{\sigma}^{(1)} +
\tfrac{1}{2}(N-1) I_{\pi}^{(1)} -
\Nc \Nf I_{\psi}^{(1)}
,
\end{equation}
where the loop functions $I_{\alpha}^{(i)}$ are defined as
\begin{align}
\label{eq:I_def} 
I_{\alpha}^{(i)} &= \Tr_q 
\left[ 
\partial_k R^A_k(q)
G_{\alpha,k}^{i}(q)
\right]\nonumber\\
&\equiv
\Tr_q 
\left[ 
\partial_k R^A_k(q)
\left(\Gamma_{\alpha,k}^{(2)}(q)+R^A_k(q)\right)^{-i}
\right],
\end{align}
with $\alpha\in\{\sigma,\pi,\psi\}$, $i>0$ and $A\in\{B,F\}$ chosen appropriately for 
bosonic and fermionic fields. The trace includes both 
a momentum integration and a trace over Dirac space for fermionic 
expressions while $G_{\alpha,k}(q)$ denotes the full Euclidean propagator at momentum 
$q$ and scale $k$ with implicit field variables replaced by their expectation 
values. In our numerical calculation we employ the three-dimensional analogue of the 
LPA-optimized regulator functions
\cite{Litim:2001up}, namely
\begin{align}
\label{eq:3dregulators}
R^B_{k}(q)&=(k^2-\vec q^{\,2})\theta(k^2-\vec q^{\,2})\,,\\
R^F_{k}(q)&=i \slashed{\vec q} \left(\sqrt{\tfrac{k^2}{\vec q^{\,2}}}-1\right)
\theta(k^2-\vec q^{\,2})\,,
\label{eq:3dregulators2}
\end{align}
for which the traces on the right hand side of the flow equations can 
be evaluated analytically which allows for a representation in terms of
bosonic and fermionic occupation numbers,
 cf.\ Appendix~\ref{app:thresholds}, a result that greatly 
alleviates the understanding of how to implement the correct analytic
continuation. 

The flow equations for the 2-point functions then read
\begin{align}
\label{eq:gamma2pion}
\partial_k \Gamma^{(2)}_{\pi,k}&=\partial_k \Gamma^{(2),B}_{\pi,k}+
\partial_k \Gamma^{(2),F}_{\pi,k}\,,\\
\label{eq:gamma2sigma}
\partial_k \Gamma^{(2)}_{\sigma,k}&=\partial_k \Gamma^{(2),B}_{\sigma,k}+
\partial_k \Gamma^{(2),F}_{\sigma,k}\,,
\end{align}
where the bosonic and the fermionic contributions are separately 
given by
\begin{align}
\label{eq:Gamma2sigmaB}
\partial_k \Gamma^{(2),B}_{\sigma,k}=&
J^B_{\sigma\sigma}(\Gamma_{\sigma\sigma\sigma}^{(0,3)})^2
+(N-1) J^B_{\pi\pi} (\Gamma_{\sigma\pi\pi}^{(0,3)})^2 \nonumber \\
&-\tfrac{1}{2}I_\sigma^{(2)}\Gamma_{\sigma\sigma\sigma\sigma}^{(0,4)}
-\tfrac{1}{2}(N-1) I_\pi^{(2)}\Gamma_{\sigma\sigma\pi\pi}^{(0,4)}\!,\\
\label{eq:Gamma2pionB}
\partial_k \Gamma^{(2),B}_{\pi,k}=&
(J^B_{\sigma\pi} +J^B_{\pi\sigma})(\Gamma_{\sigma\pi\pi}^{(0,3)})^2-\tfrac{1}{2}
I_\sigma^{(2)}\Gamma_{\sigma\sigma\pi\pi}^{(0,4)}\nonumber \\
&-\tfrac{1}{2}I_\pi^{(2)}\!\left(\Gamma_{\pi\pi\pi\pi}^{(0,4)}\hspace{-2mm}
+\!(N\!-\!2)\Gamma_{\pi\pi\tilde{\pi}\tilde{\pi}}^{(0,4)}\right),\\
\label{eq:Gamma2sigmaF}
\partial_k \Gamma^{(2),F}_{\sigma,k}=&-2\Nc \Nf J^F_{\sigma}\!,\\
\label{eq:Gamma2pionF}
\partial_k \Gamma^{(2),F}_{\pi,k}=&-2\Nc \Nf J^F_{\pi},
\end{align}
with $\pi,\tilde{\pi}\in \{\pi_1,\pi_2,\pi_3\} $ and $\pi\neq\tilde{\pi}$.
The bosonic and fermionic loop functions $J^B_{\alpha\beta}\equiv
J^B_{\alpha\beta}(p)$  and $J^F_{\alpha}\equiv J^F_{\alpha}(p)$ for
$\alpha\in\{\sigma,\pi\}$ at external momentum $p$ are defined as
\begin{align}
\label{eq:J_def}
J^B_{\alpha\beta}(p)=\:&\Tr_q
\left[ 
\partial_k R^B_{k}(q)G_{\alpha,k}(q-p)G_{\beta,k}(q)^2
\right], \\
\label{eq:J_F_def}
J^F_{\alpha}(p)=\:&\Tr_q 
\Big[ 
\partial_k R^F_{k}(q)G_{\psi,k}(q)
\Gamma_{\bar{\psi}\psi\alpha}^{(2,1)}\nonumber \\
& \left.\qquad G_{\psi,k}(q-p)\Gamma_{\bar{\psi}\psi\alpha}^{(2,1)}G_{\psi,k}(q)
\right].
\end{align}
Explicit expressions for these loop functions can also be found in 
Appendix~\ref{app:thresholds}.

\subsection{Analytic Continuation, Spectral Functions and UV 
Parameters}
\label{sec:continuation}
As mentioned in the introduction, it is generally difficult to obtain 
timelike correlation functions from Euclidean results. 
One strategy is to attempt the inverse problem of reconstructing the
corresponding spectral functions based on the known analyticity
properties of local quantum field theory from the Euclidean input
data, e.g.\ for 2-point functions. This can in principle be done with
MEM which has become a standard procedure in lattice QCD, see,
e.g., \cite{Asakawa:2000tr,Karsch:2001uw,Datta:2003ww,Ding:2012sp}. 
FRG studies have used both MEM \cite{Haas:2013hpa} and Pad\'{e} approximants
\cite{Schmidt:2011,Dupuis:2009,Sinner:2009} for this purpose as
well. Because it becomes an ill-posed problem when the input data is
too noisy and incomplete, however, here we follow a different strategy
as put forward in \cite{Strodthoff:2011tz,Kamikado2013, Kamikado2013a}
and independently in \cite{Floerchinger2012}. It is based on the
analytic continuation of the zero-component of the external momentum
in the flow equation for the 2-point function itself. We thus directly
solve flow equations for external frequencies in the complex
plane. The analytic continuation is thereby somewhat subtle in
general, and in particular at a finite RG scale $k$ where a generic
regulator would introduce additional poles in the complex
$q^0$--plane. Here we avoid this problem by using the purely spatial
$3d$-regulator functions (\ref{eq:3dregulators}) and
(\ref{eq:3dregulators2}) which do not introduce any spurious poles
during the flow. The frequency component of the $4$-momentum
transfer remains unregulated and the corresponding Matsubara sums
can be evaluated explicitly. The analytic continuation in the external
frequency then proceeds formally as in a standard one-loop calculation
in thermal field theory \cite{Das:1997gg,Lebellac:1996}. The downside
of this approach of course is that the $3d$-regulators explicitly break
the Euclidean $O(4)$ invariance. The effects of this  have been
assessed for the linear sigma model at $T=0$ and found to be
negligible for Euclidean momenta up to several hundred MeV
\cite{Kamikado2013a}. While some residual frame dependence was
observed in the spectral functions, however, their general shape and
characteristic features such as particle poles and thresholds still
remained unaffected by this breaking of Lorentz invariance in the timelike
domain. We therefore expect that it should be safe to use the $3d$
regulators here as well, especially in the rest frame of the
thermal medium.  

Explicitly, we employ the following two-step procedure for the
analytic continuation from imaginary to real time:  
Once the sum over the Matsubara frequencies in the flow equation for
the 2-point function is performed,  we first exploit the periodicity of the
bosonic and fermionic occupation numbers along the imaginary direction of
the complex energy plane, i.e.\ with respect to discrete
Euclidean external Matsubara modes $p_0=2n\pi T$,
\begin{equation}
n_{B,F}(E+\I p_0)\rightarrow n_{B,F}(E), 
\end{equation}
cf. Appendix \ref{app:thresholds}. In the second step, the retarded
2-point functions are then obtained from their Euclidean counterparts
via the analytic continuation
\begin{equation}
\Gamma^{(2),R}(\omega,\vec p)=-\lim_{\epsilon\to 0} \Gamma^{(2),E}
(p_0=-\I(\omega+\I\epsilon), \vec p).
\end{equation}
This substitution of the discrete Euclidean external $p_0$ by the
continuous real frequency $\omega $ (with small imaginary part
$\epsilon$) is done explicitly within the flow equation, before the 
integration of the RG scale $k$. Because of the one-loop structure of
the flow equations together with the unregulated Matsubara sums, the
correctness of this analytic continuation in the complex frequency
plane follows directly from the corresponding one-loop formulae for
the polarization functions in thermal field theory \cite{Das:1997gg}.
As in the one-loop examples, one therefore verifies that for the
physically correct Baym-Mermin boundary conditions \cite{Baym1961} it is crucial to
follow these two steps in this order, i.e.\ first exploit the
periodicity of the occupation numbers and then continue $p_0$ with 
$p_0\rightarrow -\I(\omega+i\epsilon)$.  

In our numerical implementation we do not take the limit $\epsilon\to 0$ but 
keep a small but finite value of $\epsilon = 1\,{\rm MeV}$. 
Moreover, we consider only the special case in which the spatial external 
momentum components vanish, $\vec p = 0$. 
Finally, the spectral function is given by the discontinuity of the
propagator and can hence be expressed in terms of  
the imaginary part of the retarded propagator as,
\begin{equation}
\label{eq:spectralreim}
\rho(\omega)=-\frac{1}{\pi}\frac{\text{Im}\,\Gamma^{(2),R}(\omega)}{\left(\text{Re}\,
\Gamma^{(2),R}(\omega)\right)^2+\left(\text{Im}\,\Gamma^{(2),R}(\omega)\right)^2}.
\end{equation}

To numerically solve the flow equation for the effective potential, Eq.~(\ref{eq:flow_pot}), 
and for the real and imaginary parts of the 2-point 
functions, Eqs.~(\ref{eq:gamma2pion}) and (\ref{eq:gamma2sigma}), we have to specify 
the model parameters and provide initial conditions. For the 
effective potential we assume the following shape in the UV,
\begin{equation}
\label{eq:pot_UV} 
U_\Lambda(\phi^{2}) =
\tfrac{1}{2}m_\Lambda^{2}\phi^{2} +
\tfrac{1}{4}\lambda_\Lambda(\phi^{2})^{2}
\,,
\end{equation}
while for the sigma and pion 2-point function we employ
\begin{eqnarray}
\label{eq:UV_sigma} 
\Gamma^{(2),R}_{\sigma,\Lambda}(\omega)&=&-\omega^2+2U_{\Lambda}'+4\phi^2 U_{\Lambda}'', \\
\label{eq:UV_pion} 
\Gamma^{(2),R}_{\pi,\Lambda}(\omega)&=&-\omega^2+2U_{\Lambda}'.
\end{eqnarray}
The model parameters are chosen to obtain phenomenologically
reasonable values for the quark and meson masses in the IR. Using the values listed 
in Table~\ref{tab:parameters}, we obtain at vanishing temperature and chemical potential
a quark mass of ${m_\psi = 299\,{\rm MeV}}$, a pion mass of ${m_\pi = 138\,{\rm MeV}}$ and a 
sigma meson mass of ${m_\sigma = 509\,{\rm MeV}}$. The chiral order parameter, to be identified with 
the pion decay constant, takes a value of ${\sigma_0 \equiv f_\pi = 93.5 \,{\rm MeV}}$. We
note here that if the temperature gets too large compared to the UV 
cutoff scale, in our case
of ${T_{{\rm max}} \approx \Lambda/2\pi \approx 170\,{\rm MeV}}$, the assumption of a temperature-independent effective 
action in the UV breaks down. If one wants to extend the accessible 
temperature range for a fixed UV cutoff, the results have to be 
supplemented by perturbative input \cite{Braun:2003ii,Herbst:2010rf,Strodthoff:2013cua}.
We checked, however, that this limitation does not change the qualitative features 
discussed in the following.

\begin{table}[t]
\centering
\begin{tabular}{C{1.3cm}|C{1.3cm}|C{1.1cm}|C{1.3cm}|C{1.3cm}}
 $\Lambda$/MeV & $m_\Lambda/\Lambda$ & $\lambda_\Lambda$ & $c/\Lambda^3$ &  $h$ \\
\hline
1000 & 0.794 & 2.00 & 0.00175 & 3.2 \\
\end{tabular}
\caption{Employed parameter values.}
\label{tab:parameters} 
\end{table}

\subsection{Numerical procedure}
\label{sec:procedure}
The flow equation for the effective potential, Eq.~(\ref{eq:flow_pot}), is solved 
by discretizing the effective potential on a grid in field space, see 
\cite{Schaefer:2004en} for details, which turns this partial
differential equation into a set of 
ordinary differential equations which are then integrated from the UV 
scale $\Lambda$ to some IR scale $k_{{\rm IR}}$. This process gets increasingly 
expensive the smaller the infrared scale at which the flow is 
stopped. Here it is chosen as $k_{{\rm IR}}= 40 \,{\rm MeV}$. 

The flow equations for the real and imaginary parts of the mesonic
2-point functions are subsequently solved at the fixed global minimum
of the effective potential in the IR, i.e.\ for $\phi=\sigma_0$. 
Typical problems that arise in this process are cutoff-effects due to
the `abrupt' start and end of the integration of the flow
equations. These cutoff effects, which can produce negative values for
the spectral functions, can  
be corrected by using extrapolations of the scale-dependent meson masses and the 
necessary derivatives of the effective potential to integrate the flow equations for the 
2-point functions to even smaller, and larger, scales. Another difficulty stems from the 
fact that the flow equations for the 2-point functions are numerically highly sensitive 
to the scale-dependence of the meson masses, $m_\sigma(k)$ and $m_\pi(k)$. At the critical point, 
for example, the sigma mass undergoes drastic changes over a small $k$-range which hampers the 
calculation of spectral functions at certain values of $\omega$, which are excluded from our discussion in the following.
A similar problem arises at higher temperatures, $T\gtrsim 50 \,{\rm MeV}$, where terms 
corresponding to the process of two pions going into a sigma meson, that are proportional to the 
derivative of bosonic occupation numbers, cf.\ Eq.~(\ref{eq:J_sym}), are no longer 
suppressed and can lead to 
negative values of the pion spectral function in a certain $\omega$ range. Where necessary, 
we therefore use an interpolation of the scale-dependent meson masses between $\Lambda$ and 
$k_{{\rm IR}}$, thus eliminating numerical fluctuations and providing a smooth shape of 
$m_\sigma(k)$ and $m_\pi(k)$, to integrate the aforementioned terms.

\section{Results}\label{sec:results}
\subsection{Phase diagram and screening masses}
\begin{figure}[t]
\includegraphics[width=\columnwidth]{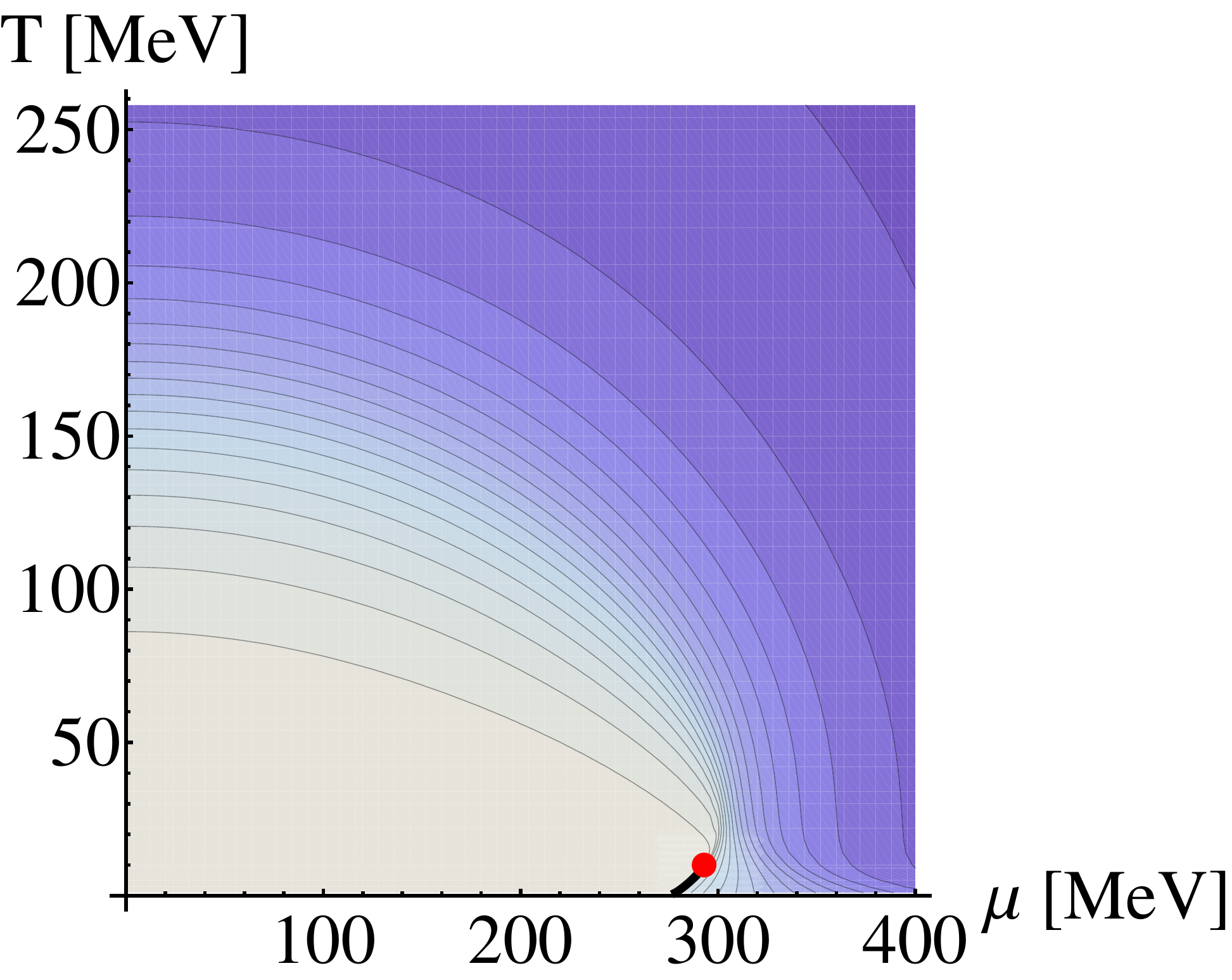}\llap{\makebox[3.5cm][l]{\raisebox{3.3cm}{\includegraphics[height=3.5cm]{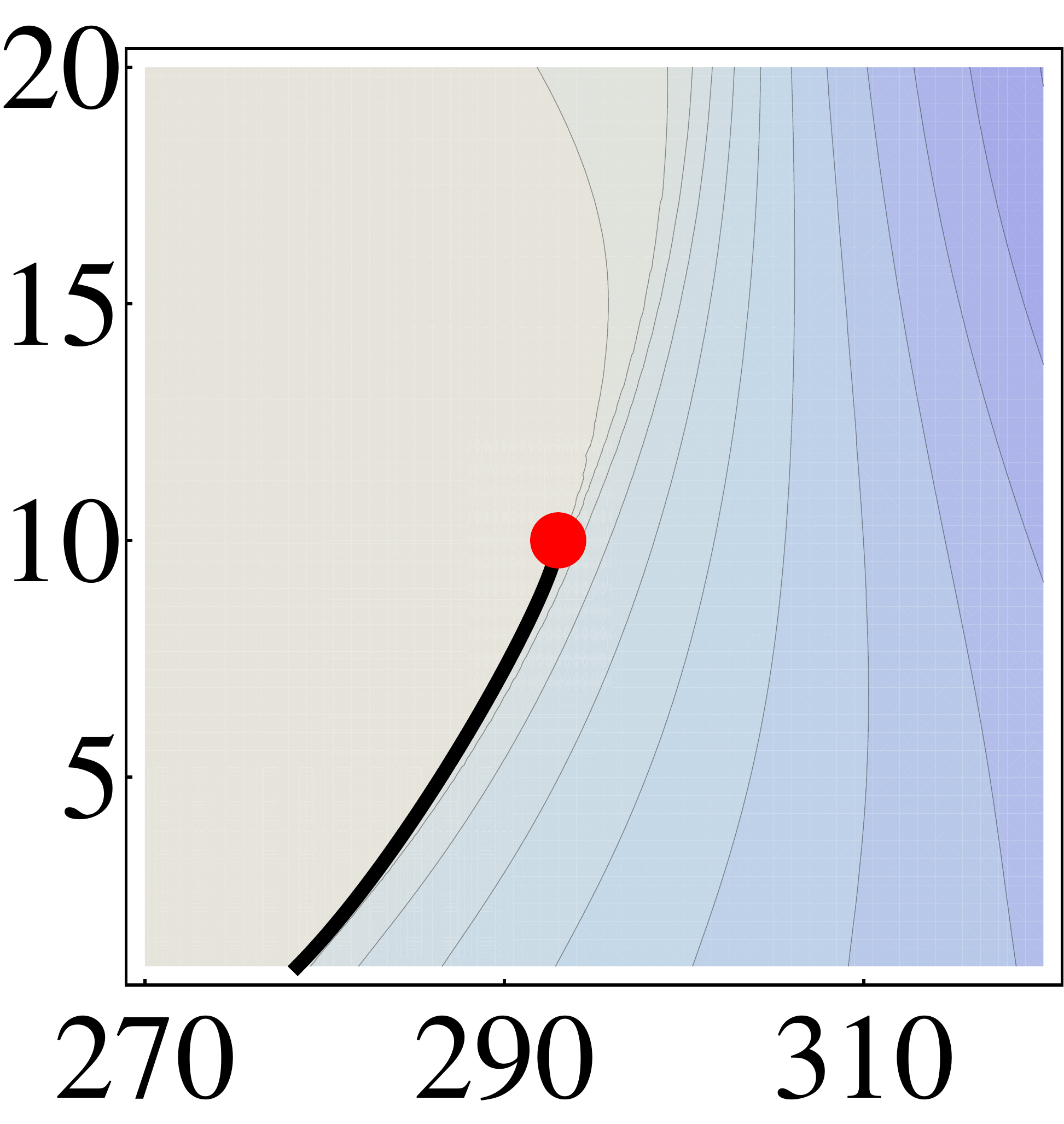}}}}
\caption{(color online) The phase diagram of the quark-meson model, obtained by using the 
parameter set in Tab. \ref{tab:parameters}, is illustrated by a contour plot of the magnitude of the chiral order 
parameter, $\sigma_0\equiv f_\pi$, vs. quark chemical potential~$\mu$ and temperature~$T$. 
The order parameter decreases towards higher $\mu$ and $T$, indicated by darker color. 
The critical endpoint is denoted by a red dot, connected to the black first-order transition 
line.}
\label{fig:phase_diagram} 
\end{figure}

We first briefly discuss the phase diagram 
obtained from the quark-meson model as it serves as input for the calculation of the 2-point 
functions here. It is shown in Fig.~\ref{fig:phase_diagram} and 
displays the typical shape \cite{Schaefer:2004en} found in quark-meson model calculations 
beyond the mean-field approximation with a critical endpoint at $\mu=293\,{\rm MeV}$ and $T=10\,{\rm MeV}$.

\begin{figure*}[t]
\includegraphics[width=\columnwidth]{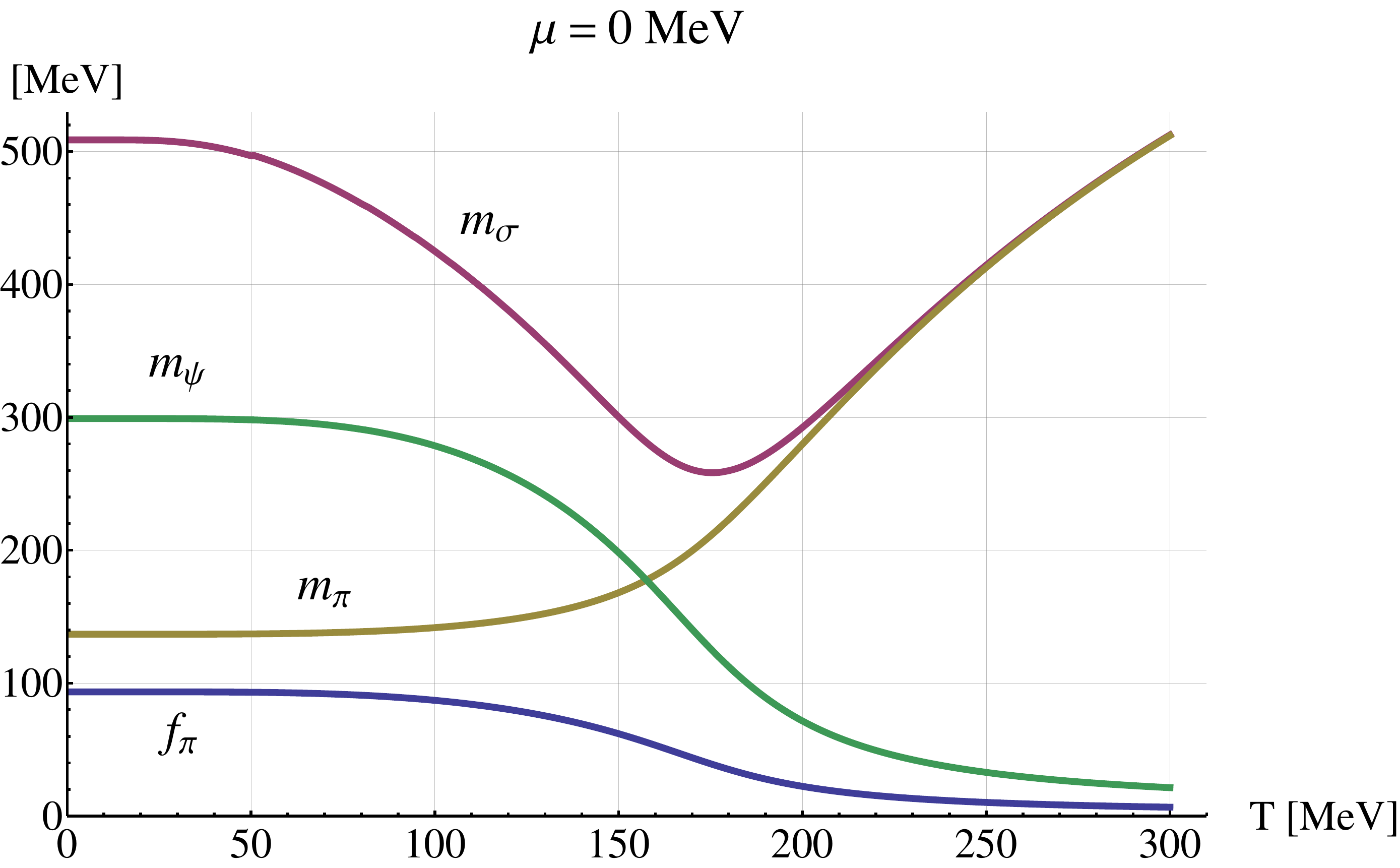}\hspace{5mm}
\includegraphics[width=\columnwidth]{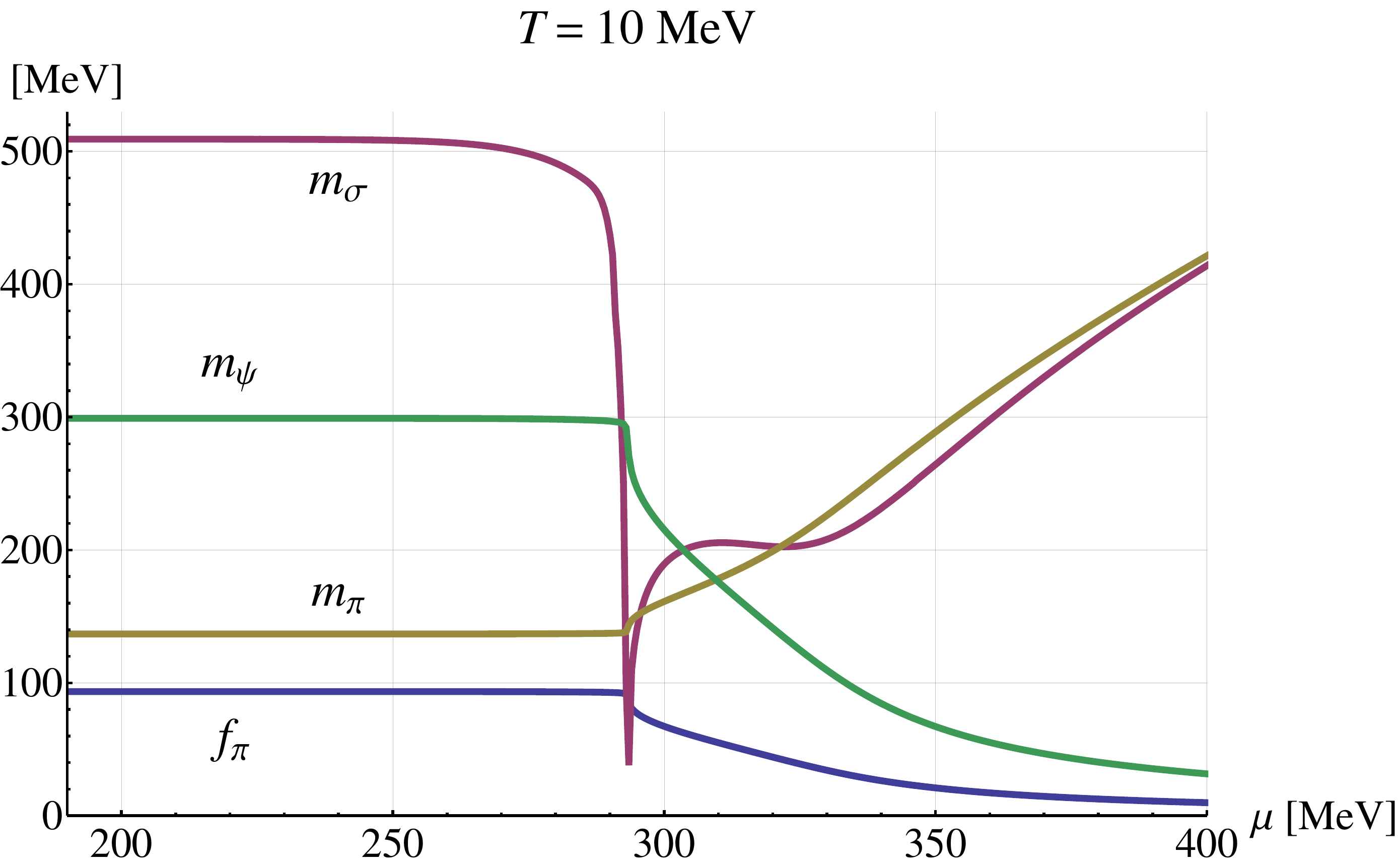}
\caption{(color online) The meson screening masses, the quark mass and the chiral order parameter, 
$\sigma_0\equiv f_\pi$, are shown vs. temperature~$T$ at $\mu=0\,{\rm MeV}$ (left panel), and 
vs. quark chemical potential~$\mu$ at $T=10\,{\rm MeV}$ (right panel).}
\label{fig:masses} 
\end{figure*}

In addition, it is also instructive to consider the 
temperature- and chemical potential-dependence of the meson screening 
masses as they will be used to identify 
thresholds in the spectral functions. The meson screening masses, 
$m_\sigma$ and $m_\pi$, can be extracted from the effective potential, cf.\ Eq.~(\ref{eq:masses}), 
and are shown in Fig.~\ref{fig:masses} together with the quark mass $m_\psi$ and the order parameter $\sigma_0$ 
as a function of temperature at 
$\mu=0\,{\rm MeV}$ (left panel) and as a function of chemical potential at $T=10\,{\rm MeV}$ (right panel).
In particular, one way of defining a pseudo-critical 
temperature for the crossover is by using the maximum of the chiral 
susceptibility $\chi_\sigma\equiv 1/m_\sigma^2$, see 
\cite{Schaefer:2006ds,Tripolt2013} and \cite{Strodthoff:2013cua} for a 
comparison of different definitions of pseudo-critical temperatures, 
which leads to value of $T_c\approx 175\, {\rm MeV}$ in our case. At 
higher temperatures the sigma and pion mass degenerate, while the order parameter drastically decreases, 
indicating the progressive restoration of chiral symmetry. 

At vanishing temperature the Silver Blaze property \cite{Cohen2003} ensures 
that the partition function and correspondingly all thermodynamic 
observables remain independent of the chemical potential $\mu$ until 
it exceeds the constituent quark mass in the 
vacuum or until it reaches a first order transition whichever occurs
first. This is approximately true here as well with the exception of 
the observed decrease in the sigma mass above $250$ MeV which is a finite  
temperature effect. At $T=10\,{\rm MeV}$ and  
$\mu\approx 293\,{\rm MeV}$, i.e.\ near the critical endpoint, the sigma 
mass suddenly further drops significantly, as expected near a second
order phase transition. At even higher values of the chemical
potential the meson masses become  degenerate again, similar to the
high temperature case. 

\subsection{Spectral functions}
\begin{figure*}
\includegraphics[width=0.95\columnwidth]{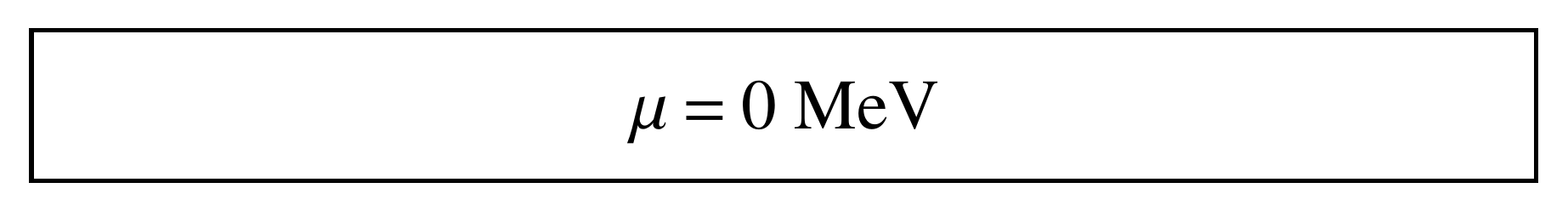}\hspace{9mm}
\includegraphics[width=0.95\columnwidth]{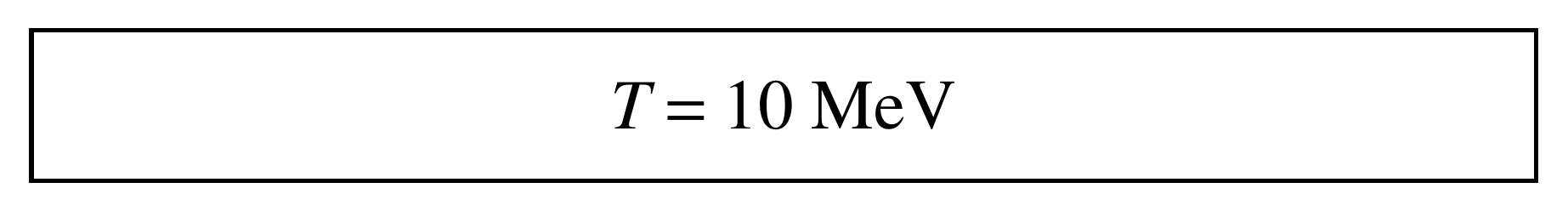}\vspace{1mm}
\includegraphics[width=\columnwidth]{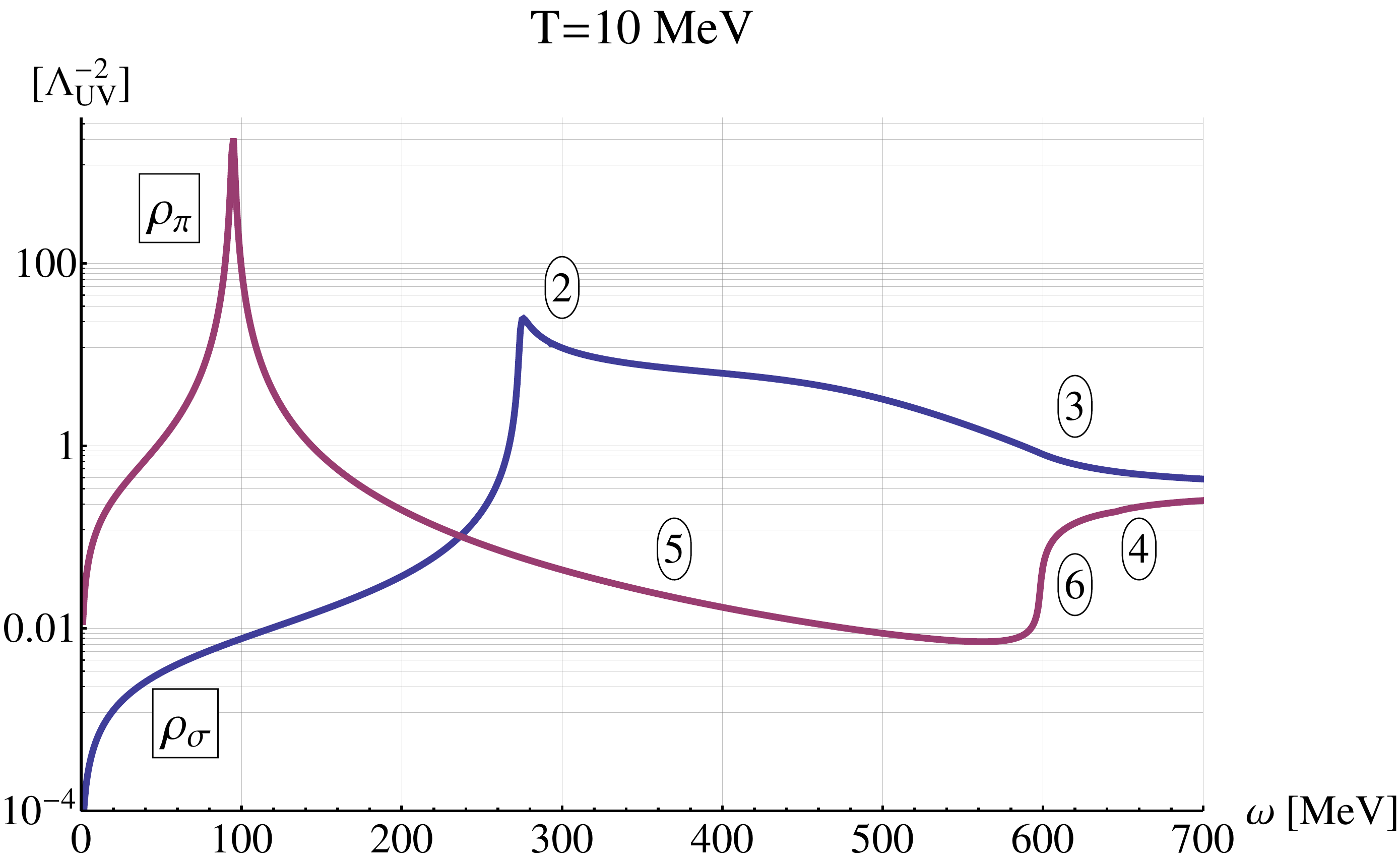}\hspace{5mm}
\includegraphics[width=\columnwidth]{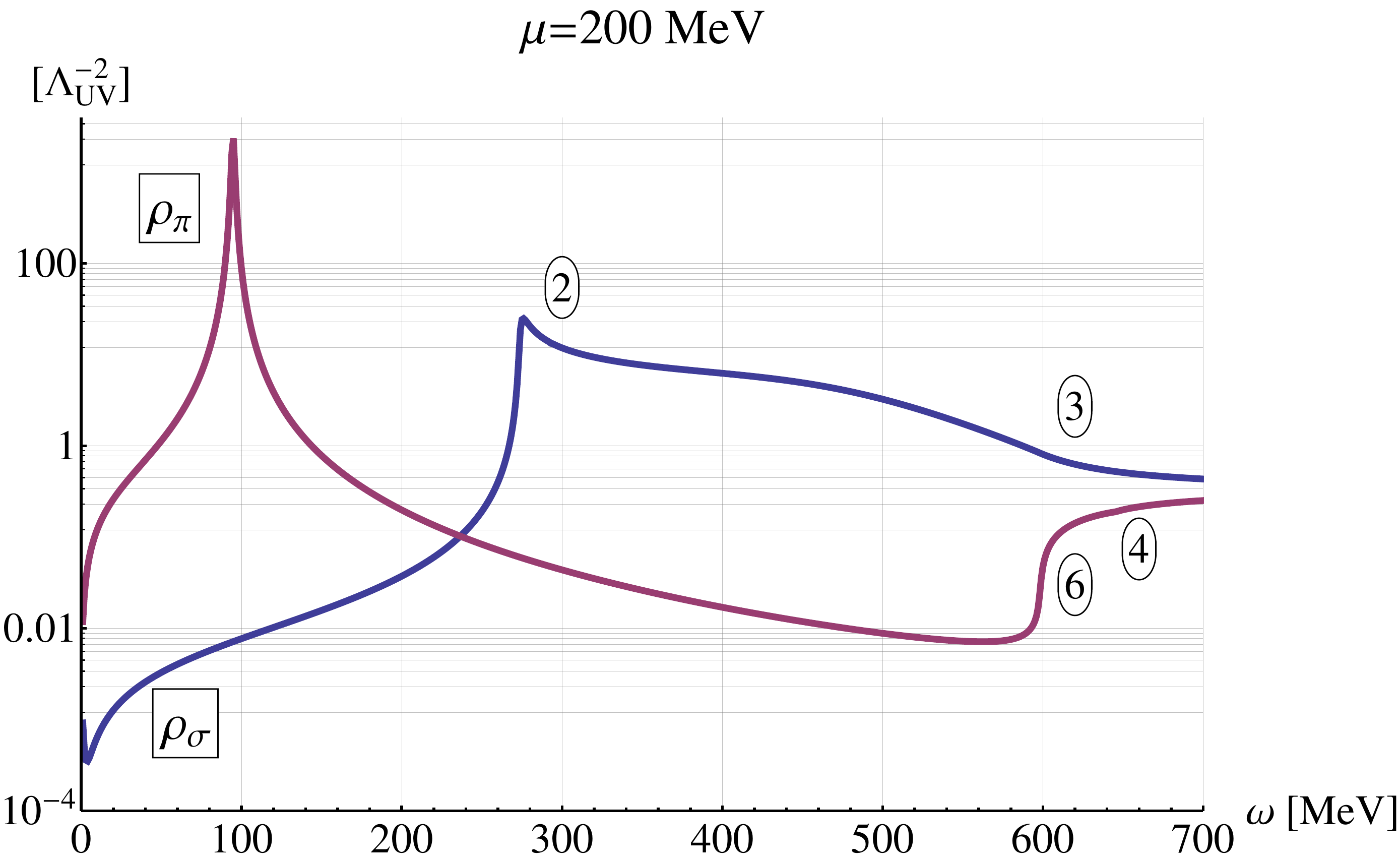}\vspace{3mm}
\includegraphics[width=\columnwidth]{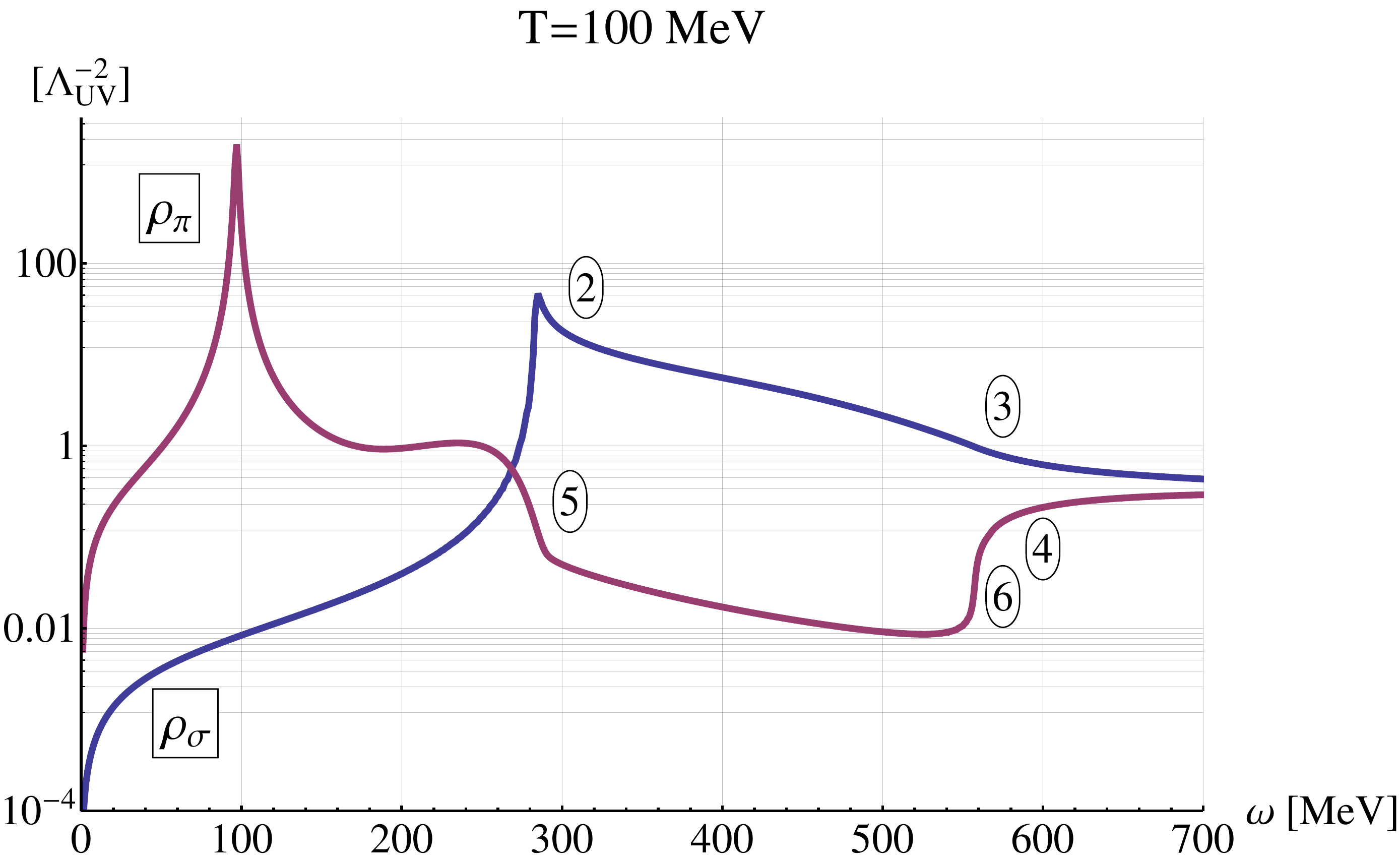}\hspace{5mm}
\includegraphics[width=\columnwidth]{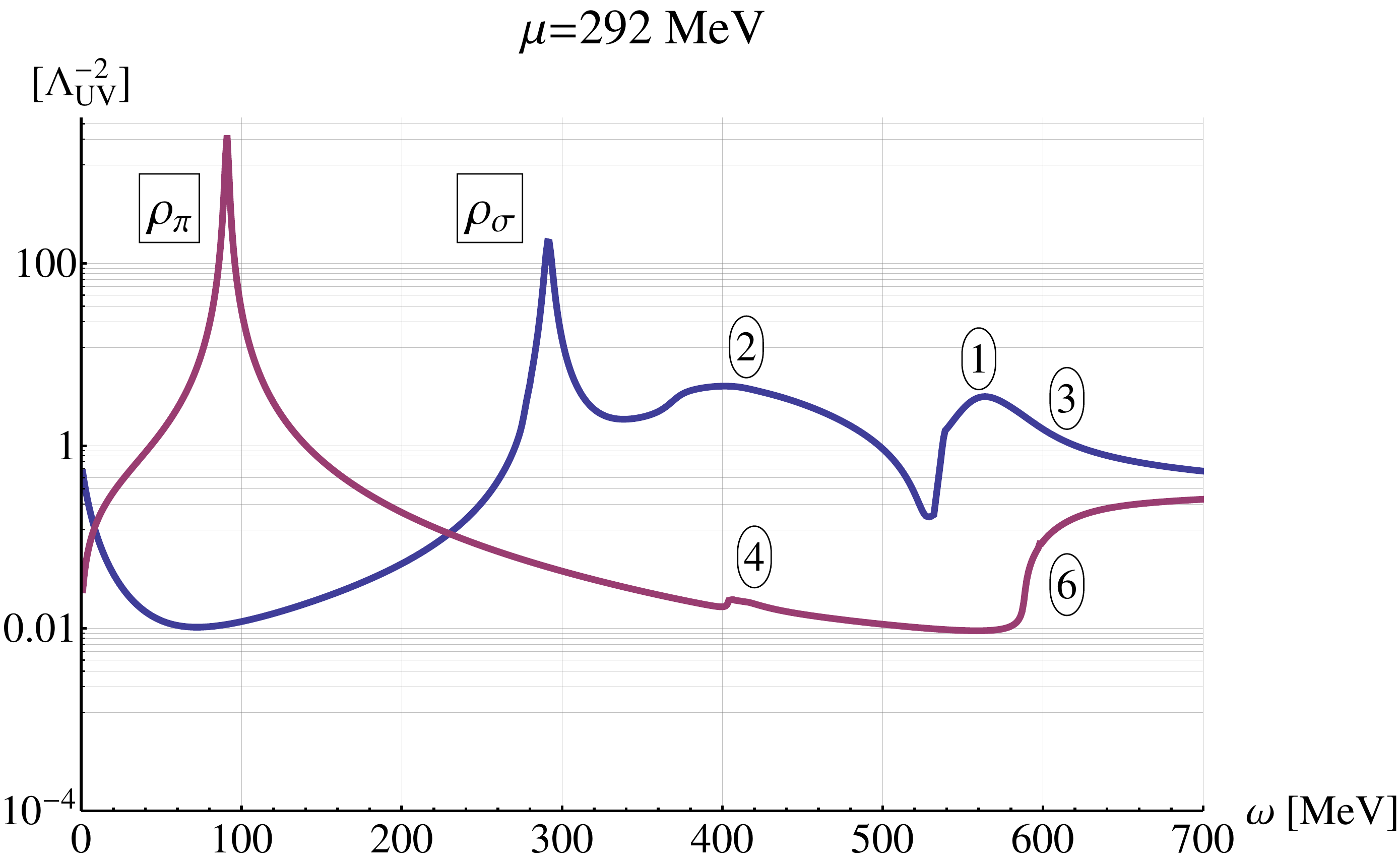}\vspace{3mm}
\includegraphics[width=\columnwidth]{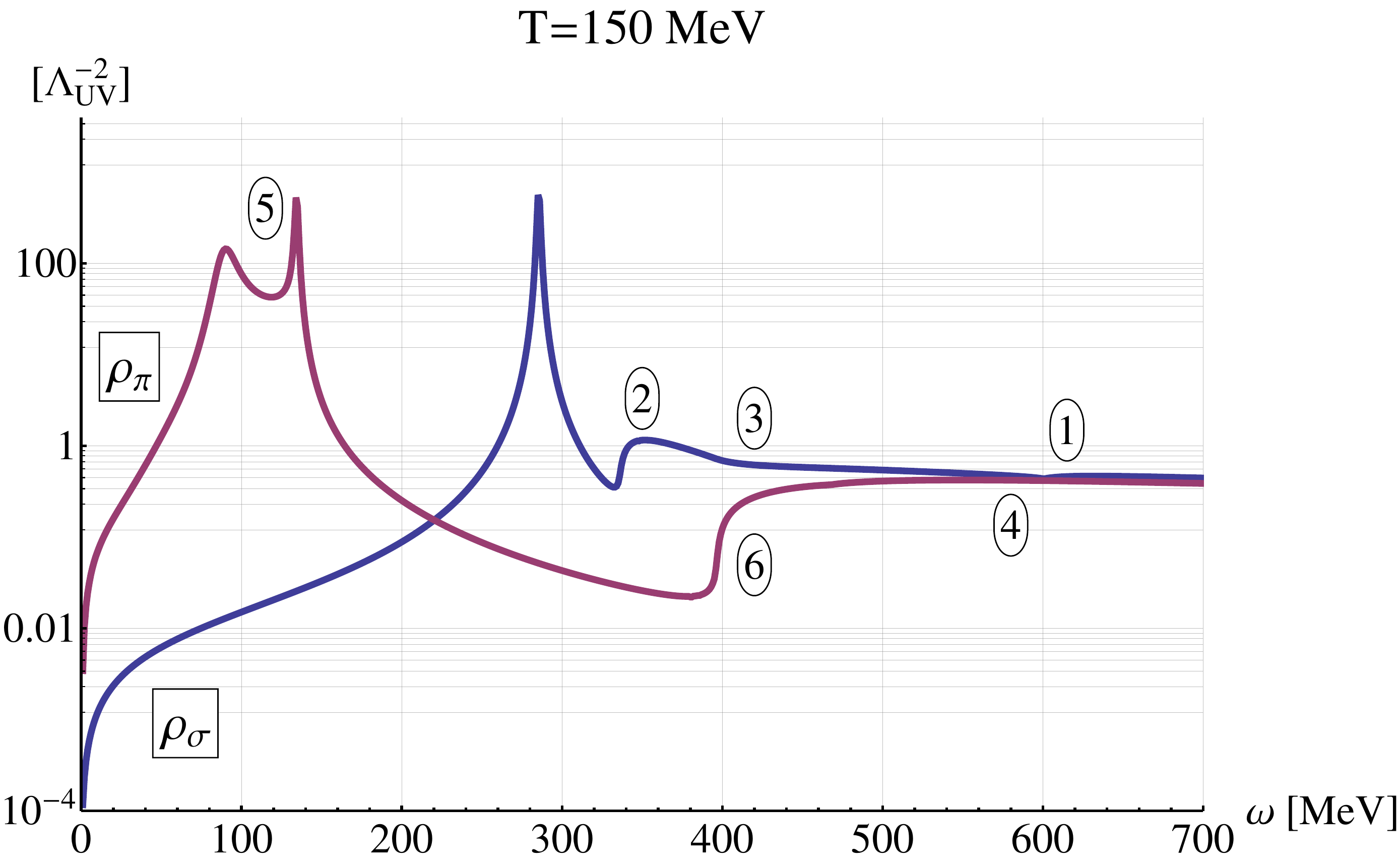}\hspace{5mm}
\includegraphics[width=\columnwidth]{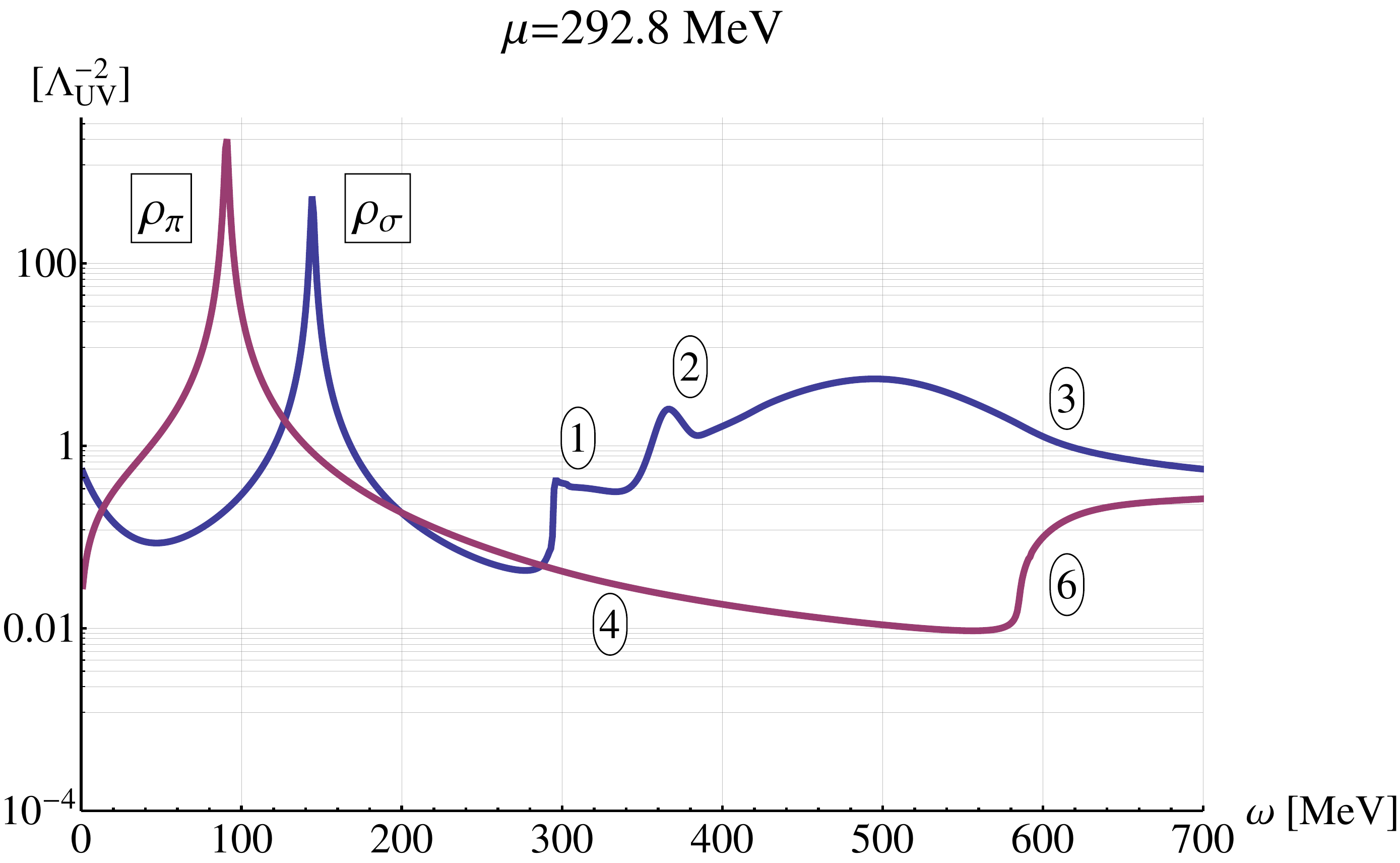}\vspace{3mm}
\includegraphics[width=0.44\columnwidth]{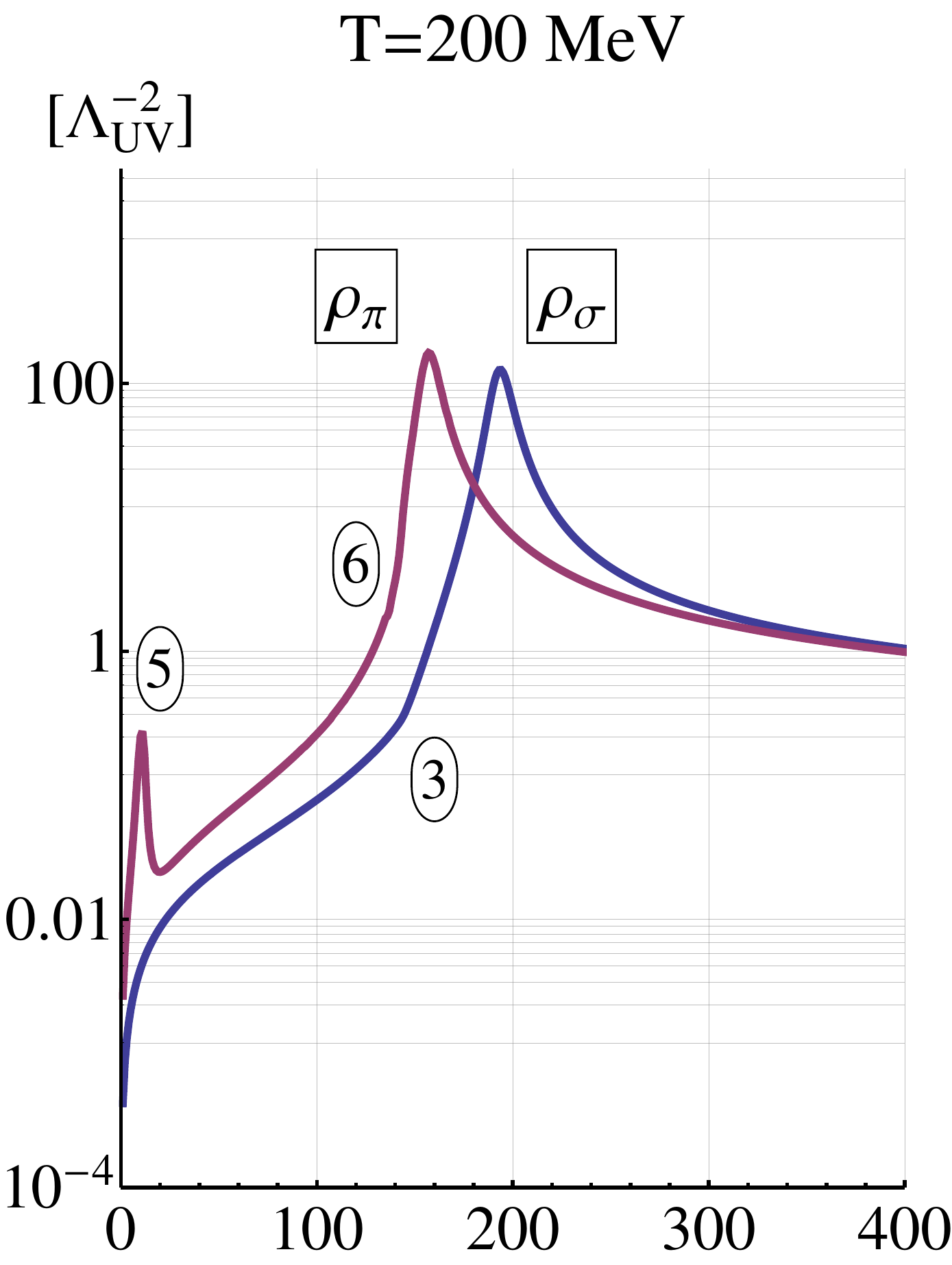}
\includegraphics[width=0.54\columnwidth]{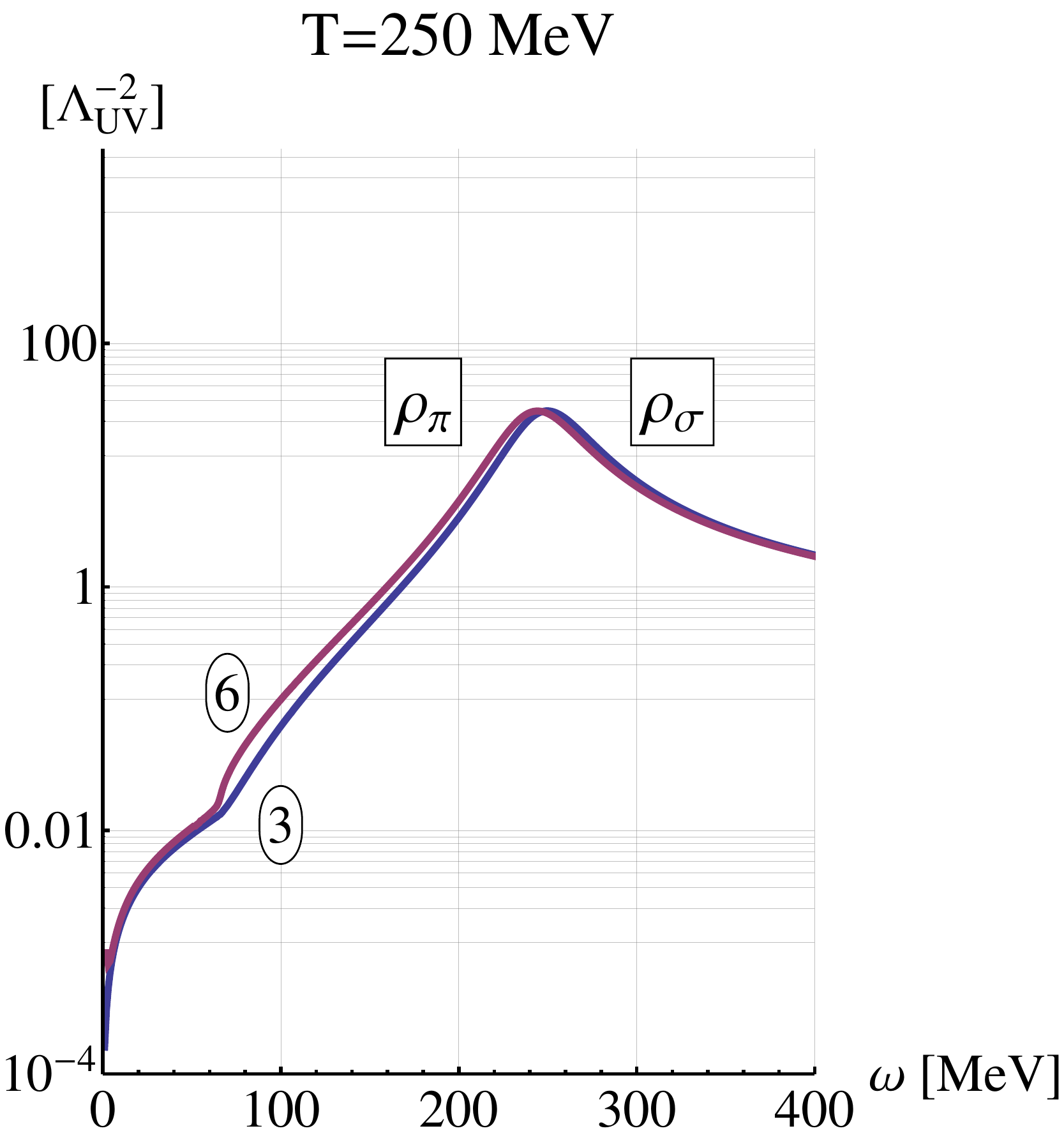}\hspace{5mm}
\includegraphics[width=0.44\columnwidth]{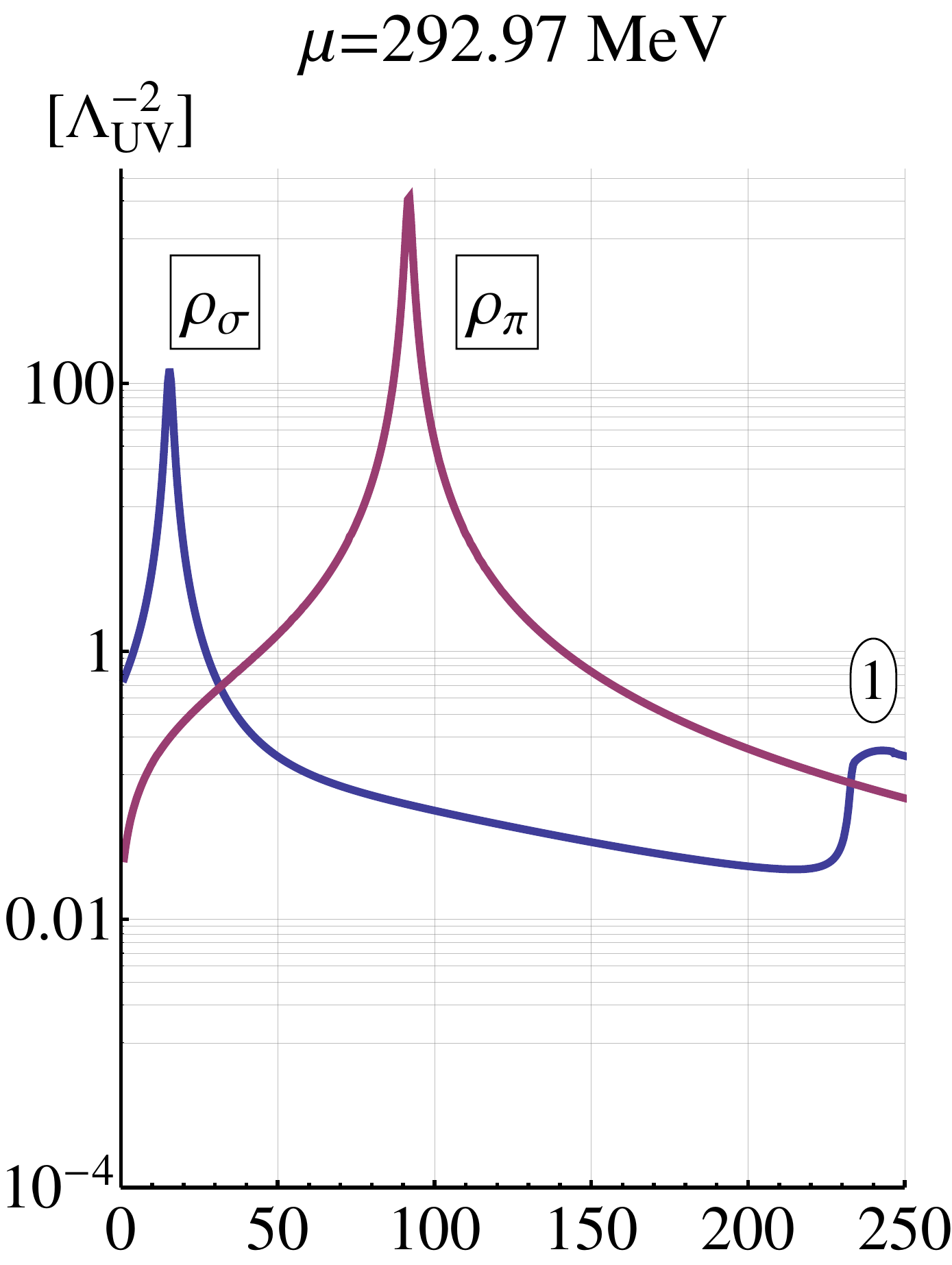}
\includegraphics[width=0.54\columnwidth]{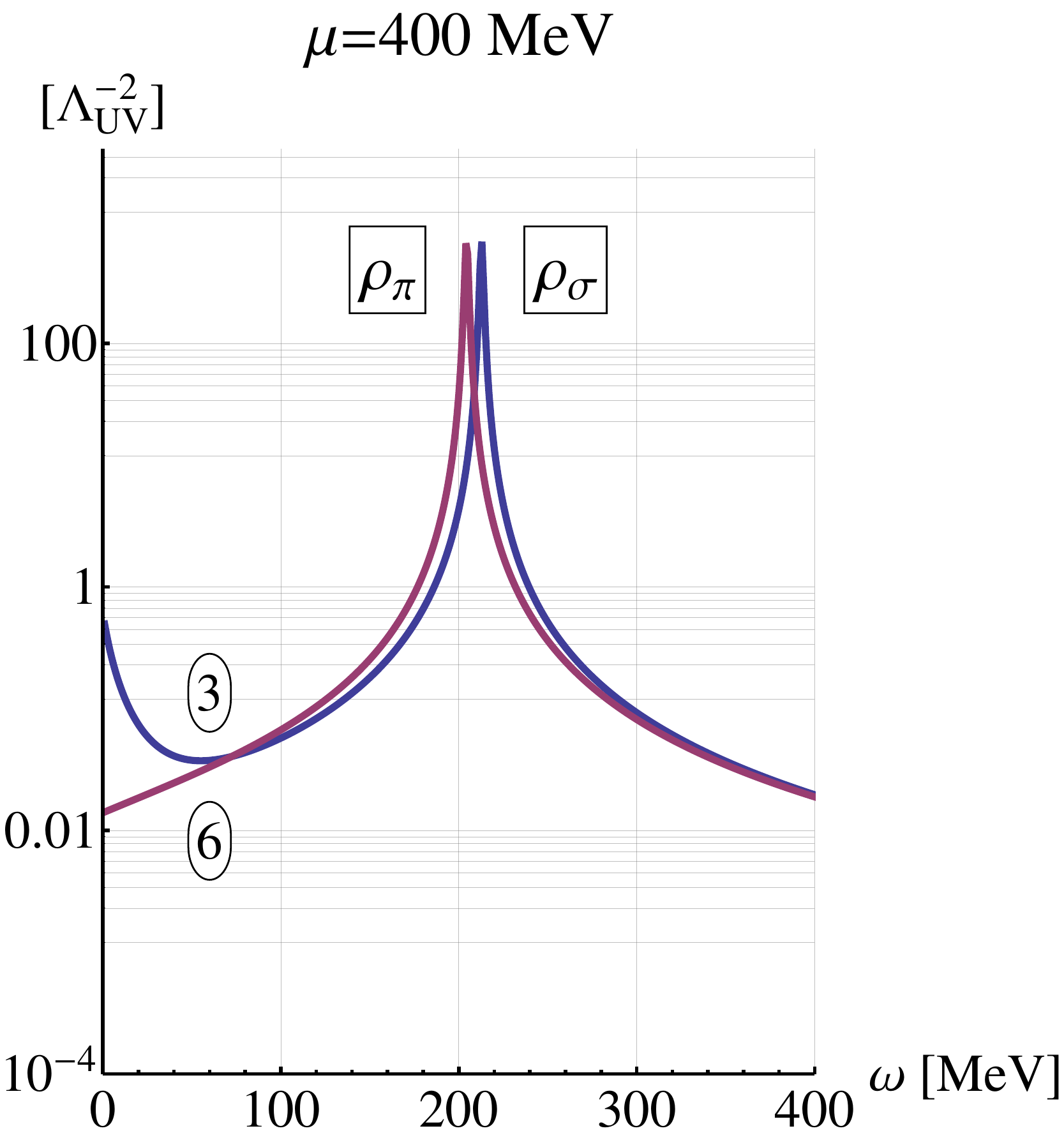}
\caption{(color online) Sigma and pion spectral function, $\rho_\sigma(\omega)$ 
and $\rho_\pi(\omega)$, are shown versus external energy $\omega$ at $\mu=0\,{\rm MeV}$ 
but different $T$ (left column) and at $T=10\,{\rm MeV}$ but different $\mu$ (right column).
Inserted numbers refer to the different processes affecting the spectral functions at the so indicated values of $\omega$. 
1:~$\sigma' \rightarrow \sigma \sigma$, 
2:~$\sigma' \rightarrow \pi \pi$, 
3:~$\sigma' \rightarrow \bar{\psi}\psi$, 
4:~$\pi' \rightarrow \sigma \pi$, 
5:~$\pi'\pi \rightarrow \sigma$, 
6:~$\pi' \rightarrow \bar{\psi}\psi$. 
See text for details.
}
\label{fig:spectralfunctions} 
\end{figure*}

We now turn to the discussion of results for the sigma and pion spectral functions, 
$\rho_\sigma(\omega)$ and $\rho_\pi(\omega)$. They are shown in Fig.~\ref{fig:spectralfunctions} 
as a function of external energy $\omega$ at different 
values of temperature and chemical potential. The inserted numbers refer to different 
processes contributing to the spectral functions via the corresponding diagrams shown in 
Fig.~\ref{fig:flow_Gamma2}. The sigma spectral function is affected by the processes 
${\sigma'\rightarrow \sigma\sigma}$ (1), ${\sigma'\rightarrow \pi\pi}$ (2) and 
${\sigma'\rightarrow \bar{\psi} \psi}$ (3), where primes denote
off-shell correlations with energy $\omega$. The relevant processes
for the pion spectral function are ${\pi'\rightarrow \sigma\pi}$ (4),
${\pi'\pi\rightarrow \sigma}$ (5) and  
${\pi'\rightarrow \bar{\psi} \psi}$ (6). In our truncation with momentum-independent vertices the mesonic tadpole diagrams only give rise to 
$\omega$-independent contributions to the spectral functions.

At $T=10\,{\rm MeV}$ and $\mu= 0\,{\rm MeV}$ the spectral functions closely 
resemble the vacuum structure already observed in previous studies, 
cf.\ \cite{Kamikado2013,Kamikado2013a}. The pion spectral function exhibits a peak close 
to $\omega =100 \,{\rm MeV}$, originating from a zero-crossing of the real part of 
$\Gamma^{(2),R}_{\pi}(\omega)$ to be identified with the pion pole mass,
cf. Appendix \ref{app:ReImGamma2} for a brief account of the real and imaginary part of the
sigma and pion 2-point-functions.
The corresponding screening mass, determined from the effective potential, 
is found to be $m_\pi = 138\,{\rm MeV}$, 
cf.\ Fig.~\ref{fig:masses}, and thus considerably larger than the pole 
mass. Discrepancies between the pole and screening masses were 
also observed in \cite{Strodthoff:2011tz} and less pronounced in purely bosonic models within similar 
truncations \cite{Svanes:2010we,Kamikado:2012bt}, and are expected to decrease with 
an increased momentum dependence of the truncation.
For external energies larger than $2m_\psi$ the decay of an (off-shell) pion into two quarks becomes 
energetically possible, giving rise to an increase of the imaginary part of 
$-\Gamma^{(2),R}_{\pi}(\omega)$ and thus also to the pion spectral function for 
$\omega\geq 2m_\psi \approx 600\,{\rm MeV}$. In addition, the process ${\pi'\rightarrow \sigma\pi}$
leads to modifications of the pion spectral function for $\omega\geq\,m_\sigma+m_\pi\approx 650\,{\rm MeV}$, 
but its contribution is about a magnitude smaller 
than that of the quark decay and therefore not visible in Fig. \ref{fig:spectralfunctions}. 
The remaining reaction affecting the pion spectral function, $\pi'\pi\rightarrow \sigma$, 
is strongly suppressed at low temperatures, since it is proportional to terms involving 
the derivative of bosonic occupation numbers, cf.\ Eq.~(\ref{eq:J_sym}). 

The dominant channel 
affecting the sigma spectral function is the decay into two pions, 
${\sigma'\rightarrow \pi\pi}$, which is possible for $\omega\geq 2m_\pi$. This process leads 
to a strong increase of the imaginary part of $\Gamma^{(2),R}_{\sigma}(\omega)$ at 
$\omega \gtrsim 275 \,{\rm MeV}$, even before the zero-crossing of the real part at 
$\omega \approx 340 \,{\rm MeV}$. At external energies larger than $2m_\psi$ the 
${\sigma'\rightarrow \bar{\psi}\psi}$ channel opens up but yields only small contributions compared to 
the decay into two pions. The process ${\sigma'\rightarrow \sigma\sigma}$ can only occur at 
external energies larger than $2m_\sigma$, i.e.\ beyond $1\, {\rm GeV}$, which are not 
shown in Fig. \ref{fig:spectralfunctions}.

When going to higher temperatures the process ${\pi'\pi\rightarrow \sigma}$, which describes 
the scattering of an off-shell pion with a pion from the heat bath resulting in the formation 
of a sigma meson, becomes less suppressed and contributes to the pion spectral function for 
$\omega\leq m_\sigma-m_\pi$. Since the difference between the meson
and pion screening masses continuously decreases with higher
temperature at $\mu=0\,{\rm  
MeV}$, cf.\ Fig.~\ref{fig:masses}, the maximum external energy up to which this process is possible is also 
shifted to smaller values, as is the corresponding bump of the pion 
spectral function, cf.\ Fig.~\ref{fig:spectralfunctions}. 

Another effect induced by the temperature dependence of 
the meson and quark masses is the emergence of a stable sigma meson at temperatures close to 
the crossover temperature, where neither the decay into two pions nor into two quarks is 
energetically possible. At $T = 150 \,{\rm MeV}$ we therefore observe a pronounced peak in the sigma 
spectral function at $\omega \approx 280 \,{\rm MeV}$, originating from a zero-crossing of 
the real part of $\Gamma^{(2),R}_{\sigma}(\omega)$. 
When increasing the temperature further, 
the quarks become the lightest degrees of freedom considered here, providing decay channels 
for both the pion and the sigma meson down to small external energies
and leading to a broadening of the pronounced peaks in the spectral functions. 

At $T = 250 \,{\rm MeV}$, the meson screening masses are nearly
degenerate and have increased to about $400\,{\rm MeV}$,  
thus shifting the threshold for decays into mesons to twice that value. Additionally, the 
quark mass has further decreased, to approximately $30\,{\rm MeV}$,
leading to a broad maximum in the almost degenerate sigma and pion
spectral functions. These effects of chiral 
symmetry restoration agree qualitatively with other studies, e.g. \cite{Hidaka2003} where
mesonic spectral functions were studied at finite temperature within an $O(N)$ model using optimized perturbation 
theory.

The right column of Fig. \ref{fig:spectralfunctions} shows the sigma and pion 
spectral function at a fixed temperature of $T=10\,{\rm MeV}$ and different 
values of the quark chemical potential, corresponding to points on a horizontal 
line in the phase diagram near the critical endpoint. We note that 
contributions to the pion spectral function arising from the process 
${\pi'\pi\rightarrow \sigma}$ are negligible at such low temperatures and are 
therefore not indicated in these figures. Over a wide range of
chemical potentials  
the spectral functions remain basically unchanged, as expected from the 
aforementioned Silver Blaze property. Between $\mu =0$ and $\mu = 200
$ MeV the results are practically identical (compare the
top left and top right panels in Fig.~\ref{fig:spectralfunctions}).   
When approaching the critical endpoint, however, especially the sigma
spectral function undergoes significant changes.  

At $\mu = 292 \,{\rm MeV}$, i.e.\ only about $1 \,{\rm MeV}$ from the
critical endpoint,  
the sigma screening mass has already dropped to about half its vacuum value, 
leading to a minimal energy for the ${\sigma'\rightarrow \sigma\sigma}$ decay of 
$\omega\geq 2\,m_\sigma\approx 540\,{\rm MeV}$. Additionally, a pronounced sigma 
peak starts to develop at $\omega \approx 290 \,{\rm MeV}$, indicating the 
formation of a stable dynamical sigma meson, cf. Appendix \ref{app:ReImGamma2}. 

Even closer to the CEP, 
i.e.\ at $\mu = 292.8 \,{\rm MeV}$, the threshold for the two-sigma
decay has decreased to $\omega \approx 290 \,{\rm MeV}$ and thus
occurs already at smaller energies  than the ${\sigma'\rightarrow
  \pi\pi}$ process. The sigma pole mass, given by the location of the
sharp peak in the sigma spectral function, has also decreased
considerably, taking a value of $m_\sigma^{p}\approx 140 \,{\rm MeV}$,
still in very good agreement with its screening mass at this point. 

At $\mu = 292.97 \,{\rm MeV}$, however, the sigma pole mass has
already decreased to $m_\sigma^{p}\approx 20 \,{\rm MeV}$, whereas the 
$\sigma'\rightarrow \sigma\sigma$ threshold, as determined by the
screening mass here, lags behind. Because the $\sigma$ screening mass
changes very rapidly close to the endpoint,
cf.\ Fig.~\ref{fig:masses}, the difference between the pole and half
the $2\sigma$-threshold is very sensitive to this remaining 
inconsistency in our present truncation and hence enhanced in this
tiny region around the endpoint. It is manifest in our approach,
however, that the pole mass and the screening mass of the sigma will both be 
zero precisely at the critical endpoint and hence identical again.    

When increasing the chemical potential further, the sigma and pion spectral 
functions become degenerate, similar to the case of high temperatures, but 
with the difference that the spectral functions exhibit sharp peaks instead 
of broad resonances, indicating stable dynamical particles. We also observe a 
small increase of the sigma spectral function at small external energies which 
originates from the different prefactors arising in the definition of the 
fermionic loop functions $J^F_{\alpha}(p_0,k)$, 
cf.\ Eqs.~(\ref{eq:JFpi})-(\ref{eq:JFsigma}), which in turn are due to the different 
Dirac structure of the quark-sigma and quark-pion vertices given in Eq.~(\ref{eq:quark_vertex}). 
This effect, however, is found to diminish when going to 
even higher values of chemical potential, as expected since the fermionic loop functions
$J^F_{\sigma}(p_0,k)$ and $J^F_{\pi}(p_0,k)$ are identical in the limit of vanishing quark mass.

\section{Summary and Outlook}\label{sec:summary}

In this work we have presented a method for calculating spectral
functions at finite temperature  
and quark chemical potential within the Functional Renormalization
Group approach, based on previous  
studies in the vacuum \cite{Kamikado2013a}. Our method involves an
analytic continuation from   
imaginary to real frequencies on the level of the flow
equations which realizes the physical Baym-Mermin boundary
conditions.
It is based on a thermodynamically consistent truncation of the 
flow equations for 2-point functions and allows  
to iteratively take into account the full momentum dependence of
2-point functions without need for  assumptions on the analytic
structure of the propagators in the future. 

Results have been obtained for mesonic spectral functions of the 
quark-meson model at finite temperature and density.
Our presentation concentrated on the temperature dependence at vanishing chemical
potential and on the vicinity of the critical point. Thereby, effects
of the following in-medium processes on the spectral functions have
been discussed:
$\pi'\rightarrow \sigma\pi$, $\pi'\pi\rightarrow \sigma$ and
$\pi'\rightarrow \bar{\psi}\psi$ for the pion 
spectral function, and $\sigma'\rightarrow \sigma\sigma$,
$\sigma'\rightarrow \pi\pi$ and $\sigma'\rightarrow \bar{\psi}\psi$ 
for the sigma spectral function. At low temperatures and small values
of chemical potential the spectral functions closely 
resemble the vacuum structure observed in previous studies \cite{Kamikado2013a}.
Near the critical endpoint in particular the decay of an off-shell
sigma meson into two 
on-shell sigma mesons results in drastic changes of the sigma spectral function. 
At high temperatures and/or large values of the chemical potential a
degeneration of the spectral functions  
was observed indicative of chiral symmetry restoration.

The results shown in this work represent a promising extension of the method 
for the calculation of spectral functions to finite temperature and
chemical potential and can be improved in several ways. Apart from the
extension to other spectral functions or the iteration procedure
described above, 
another straightforward but important extension will be the inclusion
of non-vanishing external spatial momenta. This  will then also allow
to use the method for calculations of transport coefficients such as
the shear viscosity.

\acknowledgments 
The authors thank Kazuhiko Kamikado and Jan Pawlowski for 
discussions and work on related subjects. This work was supported by
the Helmholtz International Center for FAIR within the LOEWE
initiative of the State of Hesse. L.v.S.~is furthermore supported 
by the European Commission, FP-7-PEOPLE-2009-RG, No. 249203, N.S.
by the grant ERC-AdG-290623, and \mbox{R.-A.T.} by the 
Helmholtz Research School for Quark Matter Studies, H-QM.

\appendix
\section{Real and imaginary part of the 2-point function}\label{app:ReImGamma2}
As described in Section~\ref{sec:continuation}, the spectral function can be expressed
in terms of the real and imaginary part of the retarded 2-point function, see Eq.~(\ref{eq:spectralreim}).
In this section we present results on the real and imaginary parts of the sigma and pion
2-point functions to further illustrate the behavior of the corresponding spectral functions
discussed in Section~\ref{sec:results}.

\begin{figure*}
\includegraphics[width=\columnwidth]{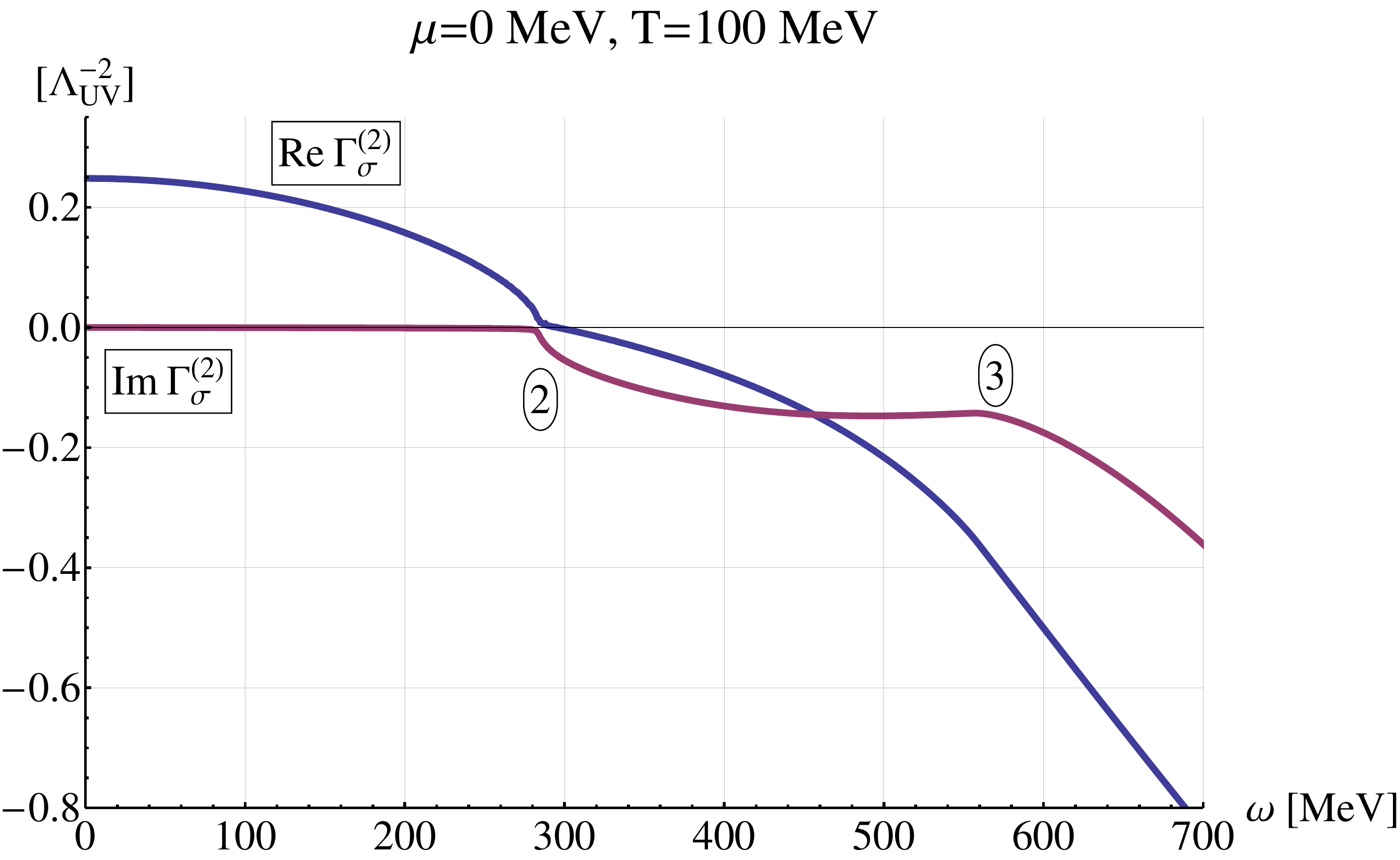}\hspace{5mm}
\includegraphics[width=\columnwidth]{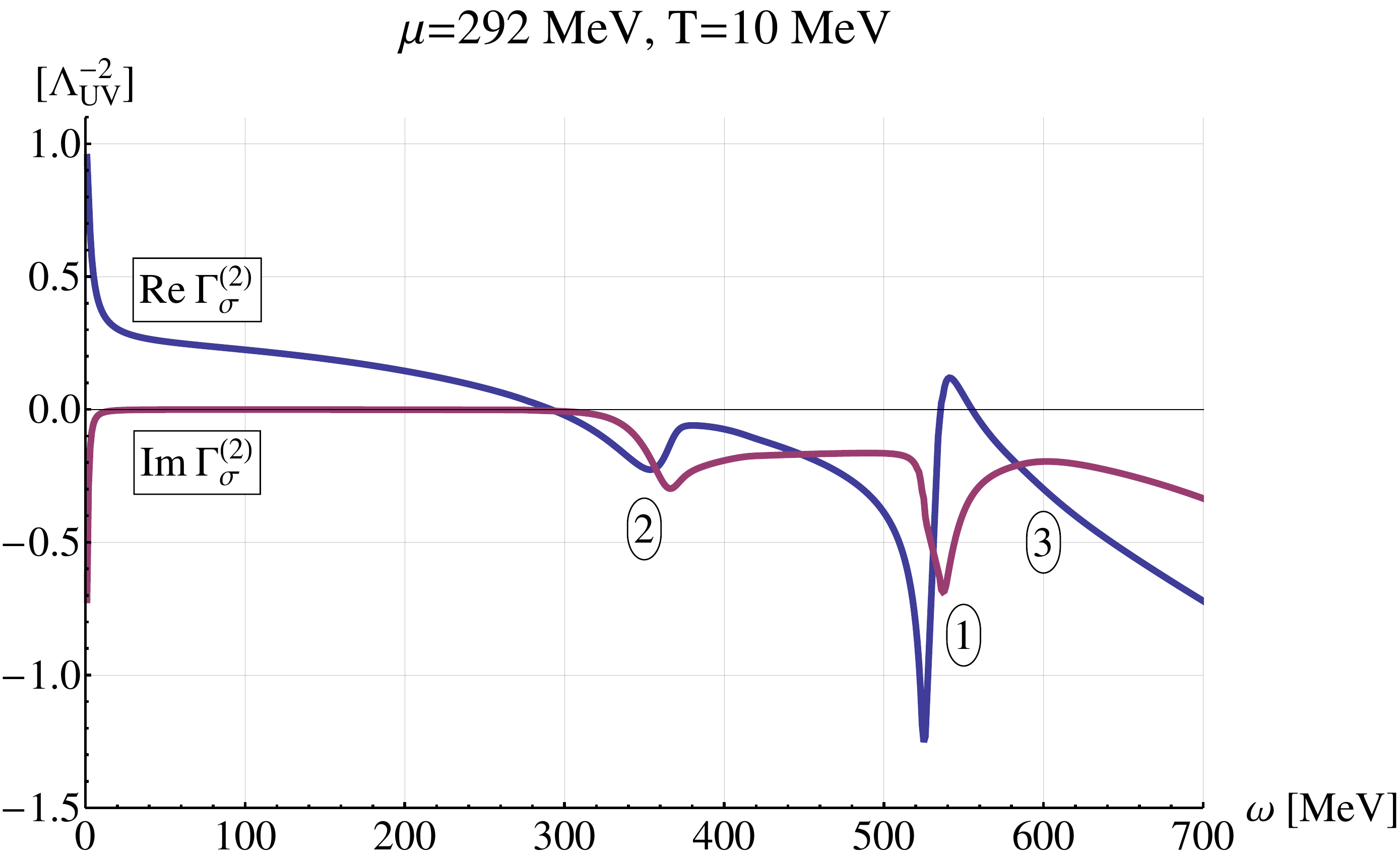}\vspace{3mm}
\includegraphics[width=\columnwidth]{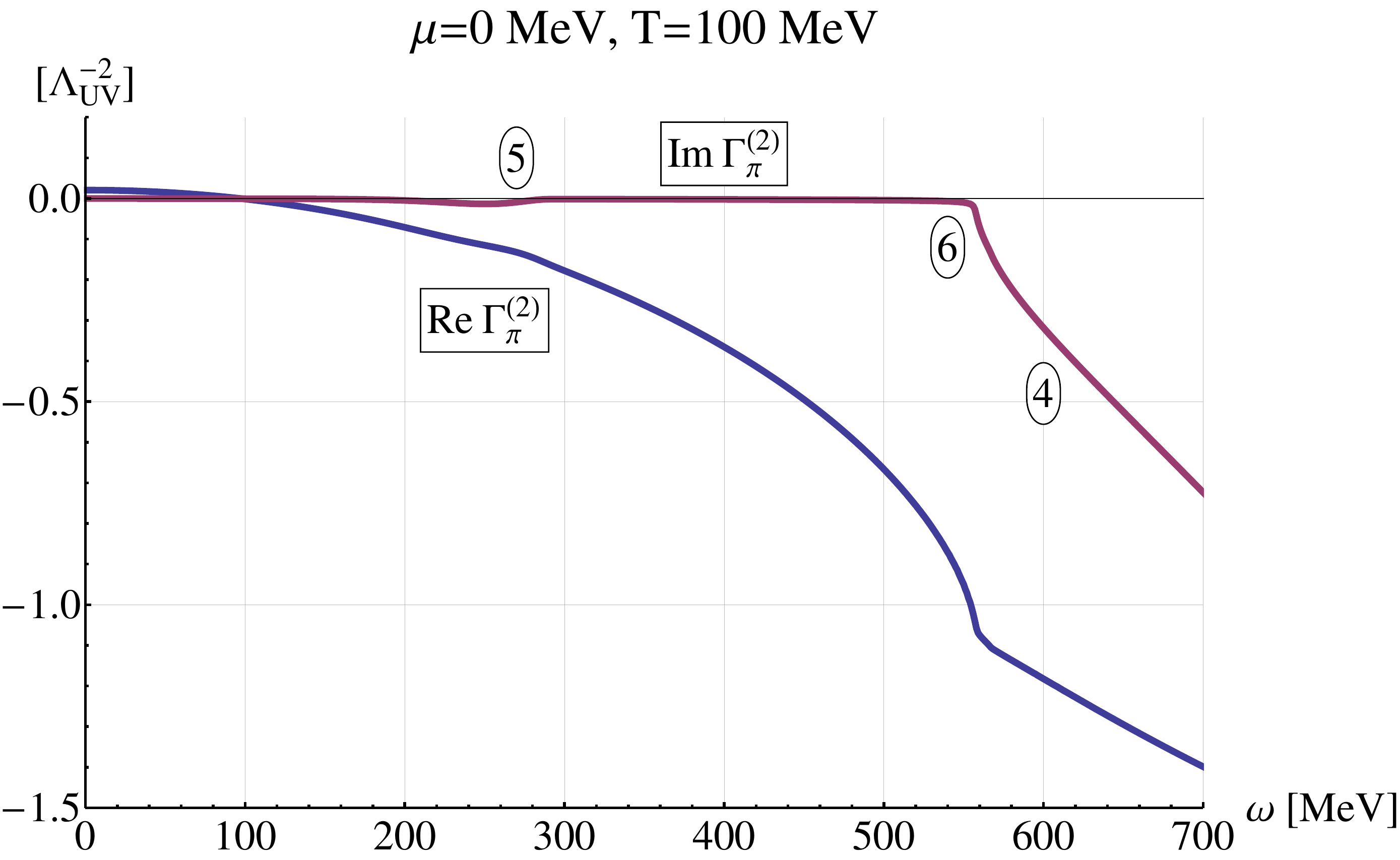}\hspace{5mm}
\includegraphics[width=\columnwidth]{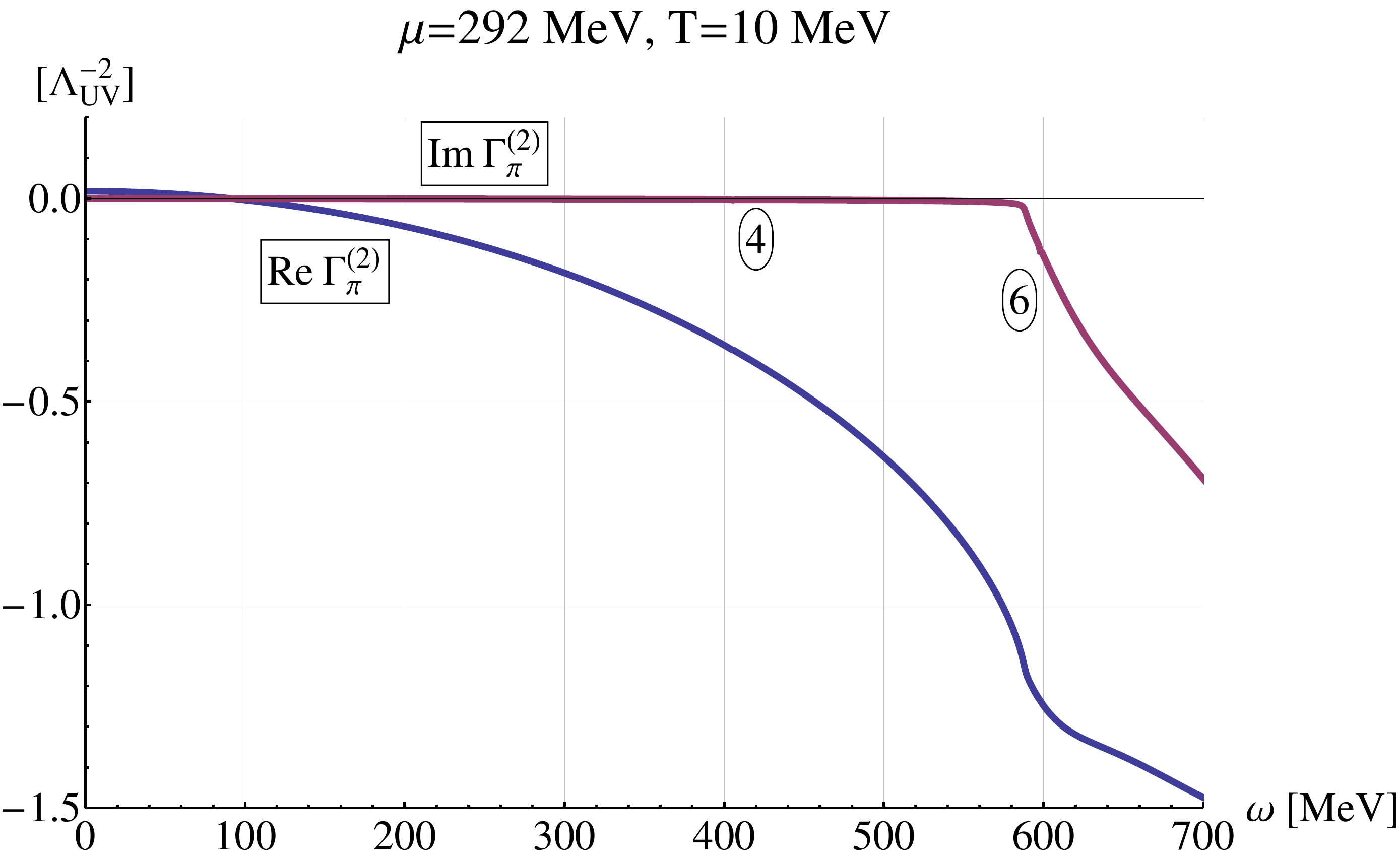}\vspace{3mm}
\caption{(color online) The real and imaginary parts of the sigma and pion two-point functions,
$\Gamma^{(2),R}_{\sigma,\pi}(\omega)$,
are shown versus external energy $\omega$ at $\mu=0\,{\rm MeV}$ and $T=100\,{\rm MeV}$ (left column)
as well as at $\mu=292\,{\rm MeV}$ and $T=10\,{\rm MeV}$ (right column), cf.\ Fig.~\ref{fig:spectralfunctions}.
Inserted numbers refer to the different processes contributing to the 2-point functions at the so indicated values of $\omega$. 
1:~$\sigma' \rightarrow \sigma \sigma$, 
2:~$\sigma' \rightarrow \pi \pi$, 
3:~$\sigma' \rightarrow \bar{\psi}\psi$, 
4:~$\pi' \rightarrow \sigma \pi$, 
5:~$\pi'\pi \rightarrow \sigma$, 
6:~$\pi' \rightarrow \bar{\psi}\psi$. 
}
\label{fig:ReandImGamma2} 
\end{figure*}

In Fig.~\ref{fig:ReandImGamma2} the real and imaginary part of the sigma and pion 2-point functions
are plotted over the external energy for two representative cases also discussed in
Section \ref{sec:results}, namely at 
$\mu=0\,{\rm MeV}$ and $T=100\,{\rm MeV}$ (left column) as well as at 
$\mu=292\,{\rm MeV}$ and $T=10\,{\rm MeV}$ (right column).

At vanishing chemical potential and a temperature of $100\,{\rm MeV}$, the real part
of the sigma 2-point function shows a zero-crossing at $\omega \approx 290\,{\rm MeV}$.
Since, however, the imaginary part takes on a non-zero value already for 
$\omega \geq 2 m_\pi\approx 280\,{\rm MeV}$, due to the decay of a sigma meson into two pions, 
the zero-crossing of the real part does not give rise to a stable sigma state.
At $\omega \geq 2 m_\psi\approx 560\,{\rm MeV}$, also the decay into a quark and an anti-quark 
contributes to the imaginary part of the sigma 2-point function.

The real part of the pion 2-point function exhibits a zero-crossing at $\omega \approx 100\,{\rm MeV}$,
giving rise to a pronounced peak in the spectral function, cf.\ Fig.~\ref{fig:spectralfunctions},
since the imaginary part does not develop non-vanishing values until external energies around $260\,{\rm MeV}$
from the $\pi'\pi \rightarrow \sigma$ process. For $\omega \geq 2m_\psi$, the quark decay channel 
and, for $\omega \geq m_\sigma+m_\pi$, also the decay into a pion and a sigma meson give rise 
to large negative values of the imaginary part of the pion 2-point function.

At a chemical potential of $292\,{\rm MeV}$ and a temperature of $10\,{\rm MeV}$, i.e.\ near
the critical endpoint, the real and imaginary part of the sigma 2-point function show large modifications from all
decay channels, in particular from the decay of an off-shell sigma state into two sigma mesons, $\sigma' \rightarrow \sigma \sigma$, 
which becomes energetically possible for $\omega \geq 2m_\sigma\approx 520\,{\rm MeV}$. 
At $\omega \approx 290\,{\rm MeV}$ the real part has a zero-crossing while the imaginary part is still close to zero,
giving rise to a pronounced peak in the spectral function, cf.\ Fig.~\ref{fig:spectralfunctions}. 

The real and imaginary part of the pion 2-point function on the other hand show only small changes compared to
their vacuum shape, cf.\ \cite{Kamikado2013a} for a discussion within the $O(N)$ model. The zero-crossing of the real part near $\omega \approx 100\,{\rm MeV}$ still gives rise to a
stable pion while at $\omega \geq 2 m_\psi\approx 580\,{\rm MeV}$ the decay into a quark-antiquark pair produces 
negative values of the imaginary part of the pion 2-point function.

\section{Threshold functions}\label{app:thresholds}
In this section we list the explicit expressions for the threshold functions appearing 
in the flow equations for the effective potential, Eq.~(\ref{eq:flow_pot}), and the bosonic 
two-point functions, Eqs.~(\ref{eq:Gamma2sigmaB})-(\ref{eq:Gamma2pionF}), using the 3d regulator 
functions from Eq.~(\ref{eq:3dregulators}).

Explicitly, the loop functions $I_\alpha^{(1)}$ and $I_\alpha^{(2)}$ are given by
\begin{eqnarray}
I_{\sigma,\pi}^{(1)} &=&
\tfrac{k^4}{6\pi^2}
\tfrac{1+2n_B(E_{\sigma,\pi})}{E_{\sigma,\pi}}
\, , \\
I_\psi^{(1)} &=&
\tfrac{k^4}{3\pi^2}
\tfrac{1-n_F(E_\psi-\mu)-n_F(E_\psi+\mu)}{E_\psi}
\, , \\
I_{\sigma,\pi}^{(2)} &=&
\tfrac{k^4}{6\pi^2}
\left(
\tfrac{1+2n_B(E_{\sigma,\pi})}{2E_{\sigma,\pi}^3}-
\tfrac{n_B'(E_{\sigma,\pi})}{E_{\sigma,\pi}^2}
\right)
\, ,
\end{eqnarray}
with the bosonic and fermionic occupation numbers
\begin{equation}
\label{eq:n_B_F} 
n_B(E)=\tfrac{1}{e^{E/T}-1},\quad
n_F(E)=\tfrac{1}{e^{E/T}+1}
\, .
\end{equation}
The effective quasi-particle energies read
\begin{equation}
\label{eq:energies}
E_\alpha=\sqrt{k^2+m_\alpha^2}, \qquad \alpha \in \{\pi,\sigma,\psi\}\,, 
\end{equation}
where the effective meson masses and the quark mass are obtained as
\begin{equation}
\label{eq:masses}
m_\pi^2=2U_k',\quad m_\sigma^2=2U_k'+4 U_k''\phi^2,\quad m_\psi^2=h^2\phi^2\,.
\end{equation}

The explicit expressions for the vertex functions appearing in Eqs.~(\ref{eq:Gamma2sigmaB})
-(\ref{eq:Gamma2pionF}) read
\begin{eqnarray}
\Gamma_{\sigma\sigma\sigma}^{(0,3)}&=&12U_k''\phi+8U_k^{(3)}\phi^3, \\
\Gamma_{\sigma\pi_1\pi_1}^{(0,3)}&=&4U_k''\phi, \\
\Gamma_{\sigma\sigma\sigma\sigma}^{(0,4)}&=&12U_k''+48U_k^{(3)}\phi^2+16U_k^{(4)}\phi^4, \\
\Gamma_{\pi_1\pi_1\pi_2\pi_2}^{(0,4)}&=&4U_k'',\\
\Gamma_{\pi_1\pi_1\pi_1\pi_1}^{(0,4)}&=&12U_k'',\\
\Gamma_{\sigma\sigma\pi_1\pi_1}^{(0,4)}&=&4U_k''+8U_k^{(3)}\phi^2.
\end{eqnarray}
The symmetrized expression for the bosonic loop functions $J^B_{\alpha\beta}(p_0)$ 
is given by Eq.~(\ref{eq:J_sym}) where $\alpha,\beta \in \{ \sigma,\pi\}$.
The fermionic loop functions $J^F_{\alpha}(p_0)$ can be cast in the form
\begin{align}
\label{eq:JFpi}
J^F_{\pi}(p_0) &= h^2\left( J^F_{0}(p_0)- m_\psi^2J^F_{1}(p_0)\right),\\
\label{eq:JFsigma}
J^F_{\sigma}(p_0) &= h^2\left( J^F_{0}(p_0)+3 m_\psi^2J^F_{1}(p_0)\right),
\end{align}
where explicit expressions for $J^F_{0}(p_0)$ and $J^F_{1}(p_0)$ are given 
by Eqs.~(\ref{eq:JF_0})-(\ref{eq:JF_1}).
\begin{widetext}
\begin{equation}
\label{eq:J_sym}
		       \begin{split}
			J^B_{\alpha\beta}(p_0)+J^B_{\beta\alpha}(p_0) =\frac{k^4}{3\pi^2}
			\Biggl( &\left[1+n_B(E_\alpha)+n_B(E_\beta)\right] 
			\frac{(E_\alpha+E_\beta)^3(E_\alpha^2+E_\alpha E_\beta+E_\beta^2) +
			(E_\alpha^3+E_\beta^3)\cdot p_0^2}{4 E_\alpha^3E_\beta^3\left(p_0^2+
			(E_\alpha+E_\beta)^2\right)^2} \\
			&-  \left[n_B(E_\alpha)-n_B(E_\beta)\right] \frac{(E_\alpha-E_\beta)^3
			(E_\alpha^2-E_\alpha E_\beta+E_\beta^2) +(E_\alpha^3-E_\beta^3)\cdot
			p_0^2}{4 E_\alpha^3E_\beta^3\left(p_0^2+(E_\alpha-E_\beta)^2\right)^2}\\
			&-\left[n_B'(E_\alpha)+n_B'(E_\beta)\right]\frac{(E_\alpha^2-E_\beta^2)^2+
			(E_\alpha^2+E_\beta^2)p_0^2}{4E_\alpha^2 E_\beta^2\left(p_0^2+(E_\alpha-
			E_\beta)^2\right)\left(p_0^2+(E_\alpha+E_\beta)^2\right)}\\
			&+\left[n_B'(E_\alpha)-n_B'(E_\beta)\right]\frac{(E_\alpha^2-E_\beta^2) 
			(E_\alpha^2+E_\beta^2+p_0^2)}{4E_\alpha^2 E_\beta^2\left(p_0^2+(E_\alpha-
			E_\beta)^2\right)\left(p_0^2+(E_\alpha+E_\beta)^2\right)}\Biggr) \\
		       \end{split}
\end{equation}
\begin{align}   
			J^F_{0}(p_0) =&  \frac{k^4}{6\pi^2}\Biggl(\left[1-n_F(E_\psi-\mu)-n_F(E_\psi+\mu)\right] 
			\frac{-4E_\psi^2(E_\psi^2+3k^2)+(5E_\psi^2-k^2)p_0^2}{E_\psi^3(4E_\psi^2+p_0^2)^2}\nonumber \\
			&-  n_F'(E_\psi-\mu) \frac{E_\psi^2-k^2+2E_\psi \I p_0}{E_\psi^2(2E_\psi+\I p_0)\I p_0}
			+  n_F'(E_\psi+\mu) \frac{E_\psi^2-k^2-2E_\psi\I p_0}{E_\psi^2(2E_\psi-\I p_0)\I p_0}\Biggr) \label{eq:JF_0}\\
			J^F_{1}(p_0) =&  \frac{k^4}{6\pi^2}\Biggl(\left[1-n_F(E_\psi-\mu)-n_F(E_\psi+\mu)\right] 
			\frac{12E_\psi^2+p_0^2}{E_\psi^3(4E_\psi^2+p_0^2)^2}\hspace{4.4cm} \nonumber\\
			&-  n_F'(E_\psi-\mu) \frac{1}{E_\psi^2(2E_\psi+\I p_0)\I p_0}
			+  n_F'(E_\psi+\mu) \frac{1}{E_\psi^2(2E_\psi-\I p_0)\I p_0}\Biggr)\label{eq:JF_1} 
\end{align}
\end{widetext}

\bibliography{qcd}

\end{document}